\documentstyle[aps,preprint,epsfig]{revtex}

\begin{document}
\tightenlines
\draft
\def\be{\begin{equation}}
\def\ee{\end{equation}}
\def\bfi{\begin{figure}}
\def\efi{\end{figure}}
\def\bea{\begin{eqnarray}}
\def\eea{\end{eqnarray}}
\title{Correlation Functions and
Fluctuation-Dissipation Relation in Driven Mixtures: an exactly
solvable model}
\author{F. Corberi$^{1,\dag}$, G. Gonnella$^{2,*}$,  E. Lippiello$^{1,\ddag}$,
 M. Zannetti$^{1,\S}$}
\address{$1$ Istituto Nazionale per la Fisica della Materia,
Unit\`a di Salerno and Dipartimento di Fisica ``E.R.Caianiello'',
Universit\`a di Salerno,
84081 Baronissi (Salerno), Italy.\\
$2$ Istituto Nazionale per la  Fisica della Materia, Unit\`a di Bari
{\rm and} Dipartimento di Fisica, Universit\`a di Bari, {\rm and}
Istituto Nazionale di Fisica Nucleare, Sezione di Bari, via
Amendola 173, 70126 Bari, Italy.}
\date{\today}
\maketitle
\begin{abstract}
The dynamics of a binary system with non conserved order parameter
under a plain shear flow with rate $\gamma $ is solved analytically 
in the large-$N$ limit. A phase transition is observed
at a critical temperature $T_c(\gamma )$. After a quench from a 
high temperature equilibrium state to a lower temperature $T$
a non-equilibrium stationary state is entered when $T>T_c(\gamma )$,
while aging dynamics characterizes the phases with $T\leq T_c(\gamma )$.
Two-time quantities are computed and the off-equilibrium generalization
of the fluctuation-dissipation theorem is provided.  
\end{abstract}

\section{Introduction}

In the field of modern statistical mechanics many efforts have been  
devoted to understand the properties of non-equilibrium.
These studies were promoted by the recognition that many relevant systems, among which
glasses and coarsening systems, are intrinsecally far from equilibrium and,
therefore, cannot be described by theories formulated as extensions
of Gibbs statistical mechanics. 
Instead, specific
concepts relative to the non-equilibrium state must be developed.
A promising subject in this field is the attempt to build
an out of equilibrium fluctuation-dissipation
theorem (FDT). Specifically, the aim of these studies is to investigate
if the relation between the (integrated) response function $\chi $ and 
the autocorrelation function $D$
is meaningful and bears some relevant informations
 regarding the non-equilibrium state.
There are several reasons for expecting this,
the most natural being induction from the equilibrium case.
In equilibrium these two quantities are linearly related by means of the
standard FDT and a fundamental quantity, the temperature of the system,
appears as the coefficient of this relation. For off-equilibrium
systems we do not have nowadays an analogous theorem, of such a wide
generality, but conjectures, indications or, sometimes, proofs exist claiming that,
in restricted contexts, the relation between $\chi $ and $D$ can be
inferred on general grounds.  
In particular, the slope of the curve $\chi (D)$ has been interpreted as
an {\it effective } temperature of the system, 
different from that of the thermal bath.

These concepts, which were originally developed in the study of turbulence
\cite{Hohenberg89} have been successively applied to the
glassy state \cite{Cugliandolo93,Cugliandolo97} and, more generally, 
to aging systems \cite{Fielding02} where 
 the non-equilibrium state is generally determined
by the change of a thermodynamic control parameter, as in a temperature
quench. 

A different non-equilibrium situation is very often considered
because of its physical relevance: The case of driven systems.
These systems are subjected to an external forcing that prevents
them to attain equilibrium.
Examples of this situation range from 
systems subjected to a  force which does not derive from a potential,
as sheared fluids or conductor wires with a potential difference applied at
their hands, to systems with externally 
imposed thermal gradients or magnetic materials under the action of an
oscillatory magnetic field and tapped materials. 
Despite many studies on this subject 
\cite{Schmittman95} much less is known
on the effect of the drive on the fluctuation-dissipation relation,
despite some progress has been done recently
\cite{Cugliandolo97,Gallavotti95}.
  
The complexity of the glassy state,  
and the difficulty of both analytical and numerical results in this field,
make the task of understanding the mechanisms whereby the FDT
can be generalized a hard question.
Therefore, in this paper we focus on binary coarsening
systems which, 
besides being physically interesting of their own, 
can be regarded 
as simple paradigms of aging behaviour.
We solve exactly the dynamics of a model for a ferromagnet
quenched from an initial  disordered state
under the action of an external driving field. This kind of
process is relevant in many physical contexts, as it  will
be discussed later. 
We consider a coarse-grained system
described by the usual double-well $\varphi^4$
hamiltonian where $\varphi$ is a scalar order parameter field. A
Langevin dynamics is considered and the order parameter is not
conserved. The driving field is  a planar shear or Couette flow.
This model, without the shear flow, is referred to in the 
literature as model A. The choice of this particular flow,
besides analytical tractability, is also
motivated by its practical relevance, which appears clearly in the
physics of complex fluids and binary mixtures, where
the behaviour of the stress as a function of the shear
rate is one of the main characterizations of fluid systems. 
Moreover, the structure of
binary and complex fluids is strongly affected by the flow and this is
relevant in many applications \cite{Larson99} and interesting
for its theoretical implications. 
In phase separation of binary mixtures, for instance, 
it was found that not only the morphology but
also the laws of dynamical scaling are affected by the
shear-induced anisotropy \cite{Onuki97,Corberi99} and
new  phenomena related to the presence of stress in segregating systems,
as logarithmic-time periodic oscillations \cite{Corberi98} and
violation of dynamical scaling \cite{braz} 
have been  observed. 

In this paper we focus on a non-conserved order parameter;
however we believe that our study could be also useful
as a preliminary step  for  understanding more complex   systems
with conserved ordered parameter. By its own the model of this
paper can describe the behaviour  of  liquid crystals undergoing
the isotropic-nematic transition  where the order parameter is not
conserved \cite{Bray94}. The phase separation properties of this
model have been studied in \cite{Cavagna00} 
where only the behaviour of  one-time correlations was considered.

The role of the non-linear term in the $\varphi^4$ hamiltonian is
crucial for quenching below and at the critical temperature.
However its presence makes the analytical study of the model
impossible unless one considers approximate theories. Among these
the so-called large-$N$ limit has played a prominent role in the
study of phase separation. The approximation takes into account
a vectorial system with an infinite number of
components $N$.
This model, which can be solved by a  self-consistent
closure, is one of the few
cases where it is possible to illustrate the out-of-equilibrium
behaviour of a coarsening system by exact calculations \cite{Bray94}.

In the case without shear
the large-$N$  approximation    captures  the essence of the phenomenon
at a semi-quantitative level \cite{Coniglio94}. 
In this work we use this approximation to calculate 
the two-time correlation function 
 showing how  the
dynamical exponents 
are  affected by the presence of the  flow. 
The response function can be also exactly computed
allowing a careful discussion of the behaviour of the fluctuation-dissipation 
relation.

The plan of the paper is the following. In Sec.~\ref{due} the model is
defined and in Sec.~\ref{duestat} its stationary 
critical properties are discussed.
In Sec.~\ref{tre} the self-consistent condition is explicitly
worked out and one-time quantities are computed.
Two-time quantities, the autocorrelation
function and the response function, are studied
in Sec.~\ref{quattro} and Sec.~\ref{cinque}. The violation of the FDT is 
considered in Sec.~\ref{sei} and  a final discussion is
presented in Sec.~\ref{sette}.
Appendixes A-D, containing some mathematical details,
complete the paper.

\section{The model} \label{due} 

We consider a system with   dynamics  described by the equations
\begin{equation}
\frac {\partial \varphi^{\alpha}(\vec x, t)} {\partial t} + \vec
\nabla \cdot (\varphi^{\alpha}(\vec x,t) \vec v) = - \Gamma \frac
{\delta {\cal H}[\vec {\varphi}]} {\delta \varphi^{\alpha}(\vec
x,t)} + \eta^{\alpha}(\vec x,t)\quad , \hskip1.5cm \alpha=1,...,N
\label{eqn2}
\end{equation}
where  $ \{\varphi^{\alpha}\}$  are the $N $ components of the
vectorial order parameter. For instance, in the case of magnetic
systems, $\vec \varphi $ is the local magnetization,
$\Gamma$ is a transport coefficient and
the gaussian white noise $\eta $, representing thermal fluctuations, 
has expectations
\begin{eqnarray}
        & & \langle  \eta^{\alpha} (\vec x, t) \rangle = 0 \nonumber \\
        & & \langle \eta^{\alpha}(\vec x, t) \eta^{\beta}
        (\vec x', t')\rangle =
        2 T \delta_{\alpha \beta}
        \Gamma \delta(\vec x -  {\vec x}\, ')
        \delta(t-t') \label{eqn3}
\end{eqnarray}
where $T$ is the temperature of the thermal bath. Without the
convective term  on the left hand side,
Eq.~(\ref{eqn2}) would be  the usual Langevin equation
which governs the purely relaxational dynamics
of a system with non-conserved order
parameter and hamiltonian $ {\cal H}[\vec {\varphi}]$. In that
case the relations (\ref{eqn3}) would assure that in thermodynamic
equilibrium at temperature $T$ the fluctuation-dissipation theorem
is verified. Here Eqs.~(\ref{eqn3}) are supposed to hold on the basis of
local equilibrium \cite{Schmittman95,Onuki79}. More precisely it is assumed that 
the non-equilibrium system can
be subdivided into cells small enough that any thermodynamic property
- which in nonequilibrium situations may depend on space -
vary slightly over one cell, but large enough that equilibrium statistical
mechanics holds within each cell \cite{Kreuzer81}. Furthermore, since
thermodynamic properties out of equilibrium may depend on time, 
the characteristic time over which a macroscopic fluctuation
dies away within one cell must be much smaller than the typical
evolution time of the system.
This implies that over intervals much smaller than
this evolution time local equilibrium is mantained
in each cell and the fluctuation-dissipation theorem
holds. This will be recovered in Sec.\ref{sei}.

The velocity $\vec v$ in Eq.~(\ref{eqn2}) is chosen to be  the
steady planar  shear flow
\begin{equation}
\vec v = \gamma y \vec e_x \label{eqn4}
\end{equation}
where $\gamma$ is the spatially homogeneous shear rate and $\vec
e_x$ is a unit vector in the flow direction. We  observe that, in
spite of the presence of a velocity field proportional to a space
coordinate, translational invariance still holds,
since a shift of $a$ in the
$y$-direction is equivalent to a galilean transformation to a new
reference frame globally moving with an added velocity $a \gamma$
in the $x$-direction. This  allows in the following to introduce
Fourier transforms and the standard definition of the structure
factor  \cite{Onuki97,Onuki79}.
The hamiltonian has  the Ginzburg-Landau form
\begin{equation}
{\cal H}[\vec \varphi] = \int_{V} d^d x \left[ \frac{1}{2} \mid
\nabla \vec  \varphi \mid^2  +  \frac{r}{2} \vec  \varphi^2 +
\frac{g}{4 N} (\vec \varphi^2 )^2 \right] \label{eqn1}
\end{equation}
where $r < 0 $, $g > 0$ and $V$ is the volume of the system. We
will be interested in processes where the system is initially in
the equilibrium ($\gamma =0$) infinite temperature state with expectations
\begin{equation}
\left \{ \begin{array}{ll} 
        < \varphi^{\alpha} (\vec x,0) >  = 0 \\
        < \varphi^{\alpha}(\vec x, 0) \varphi^{\beta} ({\vec x} \, ' ,0) > =
              \Delta_0 \delta_{\alpha \beta} \delta (\vec x- {\vec x} \, ')
	 \end{array}
\right .	
\label{ciccio} 
\end{equation}
and then is suddenly put in contact with a heat
bath at temperature $T$, and subjected to the shear flow. 
$\Delta_0$ is a constant.

In the large-N limit \cite{Bray94} the equation of motion for the
Fourier transform $\varphi^{\alpha}(\vec k,t) = \int_V d^d x
\varphi^{\alpha}(\vec x,t) \exp(i \vec k \cdot \vec x)$ takes the
linear form
\begin{equation}
\frac {\partial \varphi^{\alpha}(\vec k,t)} {\partial t} - \gamma
k_x \frac {\partial \varphi^{\alpha}(\vec k,t)} {\partial k_y} = -
\Gamma [k^2 + I(t)]\varphi^{\alpha}(\vec k, t) +
\eta^{\alpha}(\vec k, t) \label{eqn6}
\end{equation}
where \bea
  & & \langle  \eta^{\alpha} (\vec k, t) \rangle  = 0 \\ \nonumber
  & & \langle \eta^{\alpha}(\vec k, t) \eta^{\beta}(\vec k', t')\rangle
  =  2 T \delta_{\alpha \beta}
\Gamma V \delta _{\vec k ,-\vec k'} \delta(t-t') 
\eea 
and the function
\begin{equation}
I(t) =  r + \frac {g} {N} < \vec \varphi^2(\vec x, t)>
\end{equation}
has to be calculated self-consistently. The quantity 
$< \vec \varphi^2(\vec x, t)>$ can be expressed as
\begin{equation}
\frac {1}{N} < \vec \varphi^2(\vec x, t)> 
= \frac {1}{V} \sum _{\vec k} C(\vec
k,t) \label{sumsum}
\end{equation}
where the structure factor
\begin{equation}
C(\vec k, t) =  \frac {1} {NV} < \vec\varphi(\vec k, t) \cdot \vec
\varphi(- \vec k, t) >  \label{defc}
\end{equation}
is solution of the equation
\begin{equation}
\frac {\partial C(\vec k,t)} {\partial t} - \gamma k_x \frac
{\partial C(\vec k,t)} {\partial k_y} = - 2 \Gamma  [k^2 +
I(t)]C(\vec k, t)  + 2 \Gamma  T. 
\label{eqnc}
\end{equation}
The momentum sum in Eq.~(\ref{sumsum}) extends up to a phenomenological
ultraviolet cut-off $\Lambda $.

\section{Stationary properties} \label{duestat}

Letting $\partial C(\vec k,t)/\partial t=0$ in Eq.~(\ref{eqnc}),
and setting $\Gamma =1$,
the stationary form of the structure factor reads \cite{Onuki79}
\begin{equation}
C(\vec k) = 2T   \int _0 ^{\infty} e^{- 2   \int _0 ^z
{\cal K}^2(t')dt' - 2 z   \xi^{-2}_{\perp}} dz \quad , 
\label{statc}
\end{equation}
where
\begin{equation}
\vec {{\cal K}}(s) =   \vec k + \gamma s k_x \vec e_y 
\label{kstorto}
\end{equation}
and
\begin{equation}
\xi^{-2}_{\perp} = \lim_{t \rightarrow \infty} I(t) = r +   \frac
{g}{V} \sum _{\vec k} C(\vec k) . \label{xs}
\end{equation}
This quantity plays the role of a {\it transverse} 
correlation length,
because the modes of $C(\vec k)$ with $k_x=0$ 
take the usual Ornstein-Zernike form
\begin{equation}
C(k_x = 0,\vec k_{\perp}) =  \frac {T} {k_{\perp}^2 + \xi^{-2}_{\perp}}.
\label{cazz}
\end{equation}

Inserting Eq.~(\ref{statc}) into Eq.~(\ref{xs}) one has
\begin{equation}
\xi^{-2}_{\perp}  =  r +\frac {g}{V}  2T   \sum _{\vec k}
\int _0 ^{\infty} e^{- 2   \int _0 ^z {\cal K}^2(t')dt' - 2 z
  \xi^{-2}_{\perp}} dz \qquad .
\end{equation}
This equation can be  conveniently solved 
separating the $\vec k = 0 $ term
under the sum. Then,  for very large volumes
\begin{equation}
\xi^{-2}_{\perp} = r + g T P(\xi^{-2}_{\perp}) + g \frac {T} {V
\xi^{-2}_{\perp}} 
\label{sols}
\end{equation}
where
\begin{equation}
 P(\xi^{-2}_{\perp})
= 2   \int \frac {d^d k} {(2 \pi)^d} e^{-k^2/\Lambda^2}
  \int _0 ^{\infty}
e^{- 2   \int _0 ^z {\cal K}^2(t')dt' - 2 z  
\xi^{-2}_{\perp}} dz \quad .
\label{solss}
\end{equation}
The existence of a microscopic
lengthscale  proportional to $ \Lambda^{-1}$ corresponds to the
presence of a microscopic relaxation time
\be
\tau_M = (2   \Lambda^2)^{-1} 
\label{taum}
\ee
which has to be compared
with the  shear flow timescale
\be
\tau_s = \gamma^{-1}.
\label{taus}
\ee
As usually, it is assumed~\cite{Onuki97,Kreuzer81}
that  the flow does not
distort the structure of the system at microscopic level so that
$\tau_M \ll \tau_s$ and 
\begin{equation}
A \equiv \frac {2   \Lambda^2}{\gamma} \gg 1 \quad .
\label{deb}
\end{equation}
In the following we will concentrate on the particular cases
$d=2$ and $d=3$.
The function $P(x)$ is a non negative monotonically decreasing
function with the maximum value at $x=0$. It is related through
Eq.~(\ref{cucu}) to the
function $f^L(x)$ calculated in Appendix A 
and, from Eqs.~(\ref{92a},\ref{100a}),  its value at
$x=0$ is given by
\begin{equation}
P(0) =  \left\{ \begin{array}{ll} \frac {1} {4 \pi}
(\ln({\sqrt{12}A}) + \ln 2 ) & , d =2\\ 
\frac { 1 }{2
(2   \pi)^{3/2}} (\sqrt{\gamma A}  -  \Gamma^2(3/4)
 \frac {\sqrt{\gamma}} {\sqrt{2\pi}3^{1/4}}) & , d=3
\end{array} \right .
\label{cicione}
\end{equation}
where $\Gamma(3/4)$ is the Gamma function evaluated
at $3/4$. 
By graphical analysis one can easily show that Eq.(\ref{sols})
admits a solution with a finite value of $\xi _\perp ^{-2}$ for all $T$. 
However, there exists a
critical value of the temperature $T_c(\gamma)$ defined by
\begin{equation}
r + g T_c(\gamma) P(0)  = 0
\label{cicino}
\end{equation}
such that for $ T > T_c(\gamma)$ the solution is independent of
the volume, while for $T \le T_c(\gamma) $ it does depend on $V$.
From Eqs.~(\ref{cicione},\ref{cicino}) one has
\begin{equation}
T_c (\gamma)=  \left\{ \begin{array}{ll} ( 4 \pi M_0^2)/ \ln(A)\quad,
 \qquad d = 2\\ T_c(\gamma=0)
  \left [ 1 -\Gamma ^2 (3/4)   3^{-1/4} \sqrt{2\pi /A}  
\right ]^{-1} \quad, \qquad d=3
\end{array} \right .
\label{tempcrit}
\end{equation} 
with
$M_0^2 = -r / g$ and $T_c(\gamma=0) =  4M_0 ^2 \pi ^{3/2}/\Lambda$.
Notice that the effect of the flow is to increase
the value of the critical temperature with respect to the
case with $\gamma =0$ \cite{Pellicoro}, according to 
$T_c (\gamma) \sim \ln \gamma$ in $d=2 $ and  
$T_c (\gamma)-T_c(0) \propto \sqrt {\gamma}$ in $d=3$. 
In $d=2$ this produces a qualitative difference
because a  
phase-transition occurs at a finite temperature,
differently from the unsheared case. 

\begin{figure}[h]
\centering  
\vskip -40 mm
\resizebox{.87\textwidth}{!} 
{\includegraphics{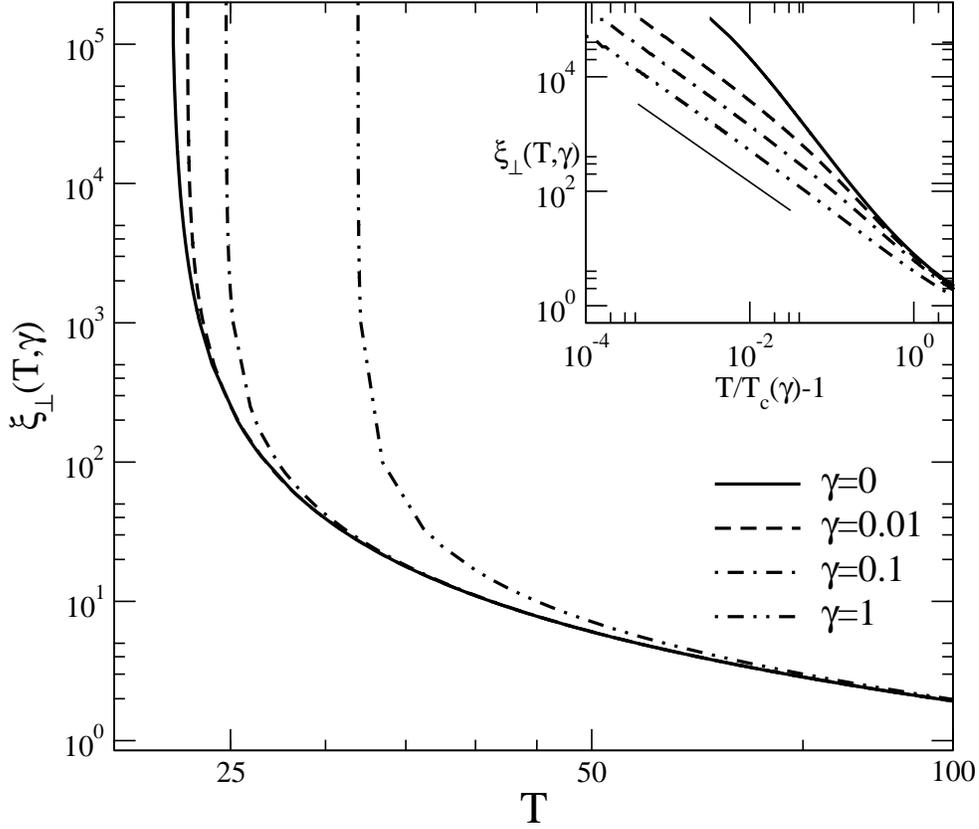} }
\caption{$\xi _\perp $ is shown against $T$ for different values of
$\gamma $ in $d=3$. The inset shows the same function plotted against 
$[T-T_c(\gamma )]/T_c(\gamma )$. The reference line below the data
curves represents the expected slope $1/2$ (Eq.~(\ref{xiperp})).}
\vskip 1.5cm
\label{figxi}
\end{figure}

Finally, concerning the behaviour of the 
transverse correlation length, above $T_c(\gamma)$
with  $[T_F - T_c(\gamma)]/T_c(\gamma)
 \ll 1$, from Eqs.~(\ref{sols},\ref{solss}) and the expression
of $f^L$ calculated in Appendix A (Eqs.~(\ref{lapf2b},\ref{lapf3b})), one finds
\be
  \xi_{\perp}^{-2} =M_0^2\left (\frac{T-T_c(\gamma)}{T_c(\gamma)}
        \right )\left (T A_3 +1/g\right )^{-1} \quad
         \mbox{$d=3$} 
\label{xiperp} 
\ee
\be
        \xi_{\perp}^{-2} |\log (\xi_{\perp}^{-2})| =
        \frac{M_0^2}{TB_2}\left (\frac{T-T_c(\gamma)}{T_c(\gamma)}
        \right )\quad
         \mbox{$d=2$}
\label{xiperp2}
\ee 
where $A_3$ and $B_2$ are defined in Appendix
A. Hence, for the exponent $\nu$ defined by $\xi_{\perp}
\sim (\frac {T - T_c(\gamma)} {T_c(\gamma)}) ^{-\nu} $ one finds
$\nu =1/2$, with logarithmic corrections
for $d = 2$. At $T_c(\gamma)$, one has $\xi_{\perp}^{-2} \sim 1/\sqrt V$
in $d=3$ and $\xi_{\perp}^{-2} |\log (\xi_{\perp}^{-2})| \sim 1/V$
for $d=2$. The behaviour of $\xi _\perp $
for $T\geq T_c(\gamma )$ is shown in Fig.\ref{figxi}.

For fixed $\gamma $, $\xi _\perp $ decreases by increasing the 
temperature. Indeed this is expected because coherence is
suppressed by thermal fluctuations. The role of $\gamma $
is more subtle. Naively one would expect that increasing
$\gamma $ produces a smaller $\xi _\perp $,
because coherent island are washed away by the flow.
However it must be also considered that $\gamma $
raises the critical temperature $T_c(\gamma )$.
Hence, for fixed $T$, larger values of $\gamma $
make $[T-T_c(\gamma )]/T_c(\gamma )$ smaller, the system gets
closer to the critical point and the coherence length
is increased. This second effect competes with the former
and produces a net increase of $\xi _\perp $ with $\gamma $.
However, by changing $\gamma $ and $T$ in order to mantain a constant
distance from criticality, namely keeping
$[T-T_c(\gamma )]/T_c(\gamma )$ fixed, the second effect is 
suppressed and one sees that, indeed, $\xi _\perp $ is
reduced by the flow (see inset of Fig.~\ref{figxi}).
For small $[T-T_c(\gamma )]/T_c(\gamma )$ the power
law behaviour~(\ref{xiperp}) is observed.
Finally, for large temperatures, $\xi _\perp $ becomes
$\gamma $-independent, because the de-coherence produced
by thermal fluctuations prevails over shear effects.

Finally, turning to the phase $T<T_c(\gamma) $, the transverse
correlation length is given by $\xi_{\perp}^{-2} = T/(M^2 V)$
where
\be
M = M_0 \sqrt{1-\frac{T}{T_c(\gamma)}} \label{magnetization} \ee
is the analogue of the spontaneous magnetization in  equilibrium.

\section{Dynamical solution} \label{tre}

We will solve the model at asymptotic times
in the case of quenching processes with initial conditions 
given by~(\ref{ciccio}).
From now on we will take the infinite volume limit.
The formal solution of Eq.~(\ref{eqnc}) is given by 
\be
        C(\vec k,t)=\frac{\Delta_0}{y(t)} 
        e^{-2\int_0^{t} du {\cal K}^2(u)} e^{-k^2/
        \Lambda ^2} +
        \frac{2T}{y(t)}  \int _{0}^{t}e^{-2\int_0^{t-z} du {\cal
        K}^2(u)} e^{-k^2/ \Lambda ^2}y(z)  dz  \label{CahnHill} \ee
where

\begin{equation}
y(t) =  \exp \left ( 2  \int _0 ^t [r  + g S(t')]dt' \right )
\qquad  \label{massren}
\end{equation}
and, from Eq.~(\ref{sumsum})
\begin{equation}
S(t) = \lim _{V\to \infty}\frac {1}{N}\langle \vec \varphi ^2(\vec x,t)\rangle=
\lim_{V\rightarrow \infty} \frac {1}{V} \sum_{\vec k} C(\vec k,t) = 
\int   \frac {d ^d k}{(2\pi)^d} e^{-k^2/\Lambda^2} C(\vec
k,t) \quad .\label{totalfl}
\end{equation}
Then, defining $f(t)$ as
\begin{equation}
f(t) \equiv  \int \frac {d^d k} {(2 \pi)^d} e^{-2   \int _0 ^t
[{\cal K}^2(z)]dz - k^2/\Lambda^2} 
\label{prior}
\end{equation}
\begin{equation}
=  (8   \pi)^{-d/2}  (t + 1/2   \Lambda^2)^{-d/2} 2
\left(4 - \frac {\gamma^2 t^4}{(t + 1/(2  \Lambda^2)^2)} +
\frac {4}{3} \frac {\gamma^2 t^3}{t + 1/(2 
\Lambda^2)}\right)^{-1/2} \label{ftd}
\end{equation}
one has 
\begin{equation}
S(t)  = \frac {1} {y(t)} \left(\Delta_0 f(t)  + 2 T    \int
_0 ^t f(t-t')y(t')dt'\right) \qquad . \label{ftf}
\end{equation}
We observe  that the correction induced by
the flow in Eq.~(\ref{ftd}) is independent on the dimensionality of the system
and that, in  absence of flow ,
the expected behaviour $f(t) \approx (8 \pi t
)^{-d/2}$ is recovered.

From Eqs.~(\ref{massren},\ref{ftf}), 
 one obtains the  integro-differential equation
\begin{equation}
\frac {d y(t)}{dt} = 2 y(t)    (r + g S(t)) =
  2 y(t)    r + 2   g
\left(\Delta_0 f(t)  + 2 T    \int _0 ^t f(t-t')y(t')
dt'\right) \label{self}
\end{equation}
that  can be solved
 introducing Laplace transforms denoted by
$f^L(s) = \int_0^{\infty} f(t) e^{-st} dt \label{lapf}$
and assuming that $ y^L(s)$, the Laplace transform of $y(t)$, is
well defined, as it will be verified  {\it a posteriori}.

Eq.~(\ref{self}) implies  that $y^L(s)$ can be expressed in terms
of $f^L(s)$ through
the equation
\begin{equation}
y^L(s) = \frac {1 + 2   \Delta_0 f^L(s)} {s - 2 r   -4 g
T  f^L(s)}. \label{lapg}
\end{equation}
The function $f^L(s)$ has been calculated in Appendix A.
Then, from  Eq.~(\ref{lapg}), inverting the Laplace transform
(see  Appendix B for details),
the  function $y(t)$ can be obtained.
For $t \gg \tau _s $, $y(t)$ behaves
as 
\vskip 1cm 
{\it d=2}
\begin{equation}
y(t) = \left\{ \begin{array}{ll}
 \frac{1}{2 B_2 T} (\frac{1}{2 }
 + \frac {M_0^2 \Delta_0 }{2 T_c(\gamma)})
\frac {1}{\log (2  \xi_\perp^{-2})} e^{-2\xi_\perp^{-2}
  t} \hskip 2cm & T   >   T_c(\gamma) \label{gtd2s} \\
\frac{1}{2 B_2 T_c}(\frac{1}{2 }
 + \frac{M_0^2 \Delta_0 }{2 T_c(\gamma)}) \frac {1}{\log t}
  \hskip 2cm & T =  T_c(\gamma)
\label{gtd2c}  \\  \frac{B_2}{M^4} (T + \Delta_0
M_0^2) t^{-2}
 \hskip 2cm & T  <   T_c(\gamma) \label{gtd2z}
\end{array} \right .
\end{equation}
\vskip 1cm {\it d=3}
\begin{equation}
y(t) = \left\{
\begin{array}{ll} \frac{1}{2}
 e^{-2\xi_\perp^{-2}   t }\frac{1 + M_0^2 \Delta_0  /T_c(\gamma)}
{1 + 4 T_c(\gamma) A_3  } \hskip 2cm & T  >  T_c(\gamma) \\
\label{gtd3s}
 \frac{1}{2} \frac{1 + M_0^2 \Delta_0  /T_c} {1 + 4
T_c(\gamma) A_3  }
 \hskip 2cm & T  = T_c(\gamma) \\
\label{gtd3c}  \frac{3}{4} \frac{1}{\sqrt{\pi}}\frac{B_3}{M^4} (
T + \Delta_0 M_0^2) t^{-5/2}
 \hskip 1cm  & T  <  T_c(\gamma)
\label{gtd3z}
\end{array} \right .
\end{equation}
where $M$ is given in Eq.~(\ref{magnetization}),
 $\xi_\perp$ and $T_c(\gamma)$ are given in
Eqs.(\ref{tempcrit},\ref{xiperp},\ref{xiperp2})
and the parameters $A_3, B_2, B_3$ are defined
in Appendix A.
Eqs.~(\ref{gtd2c},\ref{gtd3c}) hold for $t> \xi _\perp ^{2}$ in
the high temperature phase $T>T_c(\gamma )$ and for $t \gg \tau _s$
when $T\leq T_c(\gamma )$.   

\section{Autocorrelation function} \label{quattro}

In this Section we analyze the asymptotic behaviour of the
autocorrelation function
\begin{equation}
D(t,t') = \int \frac {d^d k} {(2 \pi)^d} \int \frac {d^d k'} {(2
\pi)^d} {\cal D}(\vec k,t;\vec k',t') e^{-k^2/2\Lambda^2}
e^{-k'^2/2\Lambda^2}  \label{2.1}
\end{equation}
where the correlator
\begin{equation}
{\cal D}(\vec k,t; \vec k',t') = <\varphi(\vec k, t) \varphi (\vec
k',t')> \label{defd}
\end{equation}
satisfies the equation
\begin{equation}
\frac {\partial {\cal D}(\vec k,t; \vec k',t') } {\partial t} - \gamma
k_x \frac {\partial {\cal D}(\vec k,t; \vec k',t') } {\partial k_y} = -
  [k^2 + g S(t) +r]{\cal D}(\vec k,t; \vec k',t')  \label{eqnd}
\end{equation}
with initial condition ${\cal D}(\vec k,t'; \vec k',t') = 
C(\vec k', t') \delta(\vec k + \vec k')$ and $t \ge t'$.

The function $D(t,t')$ is calculated in Appendix D 
using Eqs.~(\ref{gtd2c},\ref{gtd3c}) for $y(t)$. 
Here we  discuss  the results
in the three cases $T>T_c(\gamma )$, $T=T_c(\gamma )$, and $T<T_c(\gamma )$.

\vskip 1 cm
\noindent $\mathbf {T>T_c(\gamma )}$

In the regime $t' >\xi^{2}_{\perp} $
Eqs.(\ref{ddhat},\ref{riass1})
show that the correlation
function becomes the TTI quantity 
\be
D(t,t')\simeq D_{st}(\tau , \xi _\perp) = 
\frac {4T}{(8\pi )^{d/2}} \int _{\tau/2}^\infty
    e^{-2 \xi_\perp^{-2}   y}
    (y+1/2\Lambda^2)^{-d/2} \frac {1}{\sqrt{4 +\frac{1}{3}\gamma ^2 y^2
    + \frac{\gamma ^2}{2} \tau ^2 \left (1-\frac{\tau^2}{8y^2} \right
    ) }}dy
\label{dddhat}
\ee
with  $\tau =t-t'$. This implies that $\xi _\perp ^2$ is the 
characteristic relaxation time of fluctuations above the critical
temperature.

\vskip 1 cm
\noindent $\mathbf {T=T_c(\gamma )}$

At the critical temperature, due to  Eqs.(\ref{ddhattc},\ref{riass2}), 
$D(t,t') $ tends, for $t' \gg \tau _s $ (Eq.~(\ref{taus})), to the form
$D_{st}(\tau,\infty )$, namely
\be
D(t,t')\simeq D_{st}(\tau,\infty) = 
\frac {4T}{(8\pi )^{d/2}} \int _{\tau/2}^\infty
    (y+1/2\Lambda^2)^{-d/2} \frac {1}{\sqrt{4 +\frac{1}{3}\gamma ^2 y^2
    + \frac{\gamma ^2}{2} \tau ^2 \left (1-\frac{\tau^2}{8y^2} \right
    ) }}dy.
\label{ddddhat}
\ee
\vskip 1 cm
Then one has TTI again but, for large $\tau $, $D_{st}(\tau,\infty)$ 
decays as a power
law $D_{st}(\tau ,\infty)\propto \tau ^{-d/2}$, implying the absence
of a characteristic relaxation time, as usual at criticality.
Notice that the shear has the effect of increasing  the exponent
of the power law decay with respect to the value $-(d-2)/2$
of the undriven case. In other terms, the flow tends to decorrelate faster
the system.

\vskip 1cm
\noindent $\mathbf {T<T_c(\gamma )}$

In this case there are two time regimes of interest:

{\it i}) short time separation:
$t'\to \infty \quad , \quad \frac{\tau } {t'} \to 0 \quad$ (Quasi-stationary regime)

{\it ii}) large time separation:
$t'\to \infty \quad , \quad \frac{\tau } {t'} \to \infty \quad$ (Aging regime).

\noindent

In the time sector
{\it i}), from
Eqs.(\ref{riass3}) one has 
\be
D(t,t')= M^2 + D_{st}(\tau , \infty)
\label{eqstatio}
\ee 
where $D_{st}(\tau , \infty)$ is the same time translational invariant quantity
found at $T=T_c(\gamma )$, Eq.~(\ref{ddddhat}).

On the other hand, in the limit {\it ii}), 
from Eq.~(\ref{riass4}), one gets
\be
     D(t,t')=D_{ag}(t/t')=
    M^2 \left ( \frac{t'}{t} \right )^{\frac{d+2}{4} }
    (1+t'/t)^{-\frac{d+2}{2}}
    \frac {2^\frac{d+2}{2}}{ \sqrt{ 4 \frac {2-(1-t'/t)^3}{1+t'/t}- 3
    \frac {(2-(1-t'/t)^2)^2}{(1+t'/t)^2} }} .
\label{eqaging} 
\ee 
Notice the dependence of $D_{ag}$ on the ratio $t/t'$ alone,
as usual in slowly relaxing aging systems.
Furthermore,
for $ t/t' \gg 1 $, a generalization of the
Fisher-Huse exponent $\lambda$ defined by
\be
D(t,t')\propto (t'/t)^{\lambda } \label{2.21} 
\ee 
gives
\be
\lambda = \frac{d+2}{4} . 
\label{2.21b} 
\ee 

The structure of the autocorrelation function provided in 
Eqs.~(\ref{eqaging},\ref{eqstatio}) for large $t'$ allows its
splitting into the sum of two distinct contributions
\be
D(t,t')=D_{st}(\tau ,\infty)+D_{ag}(t/t').
\label{ssplitt}
\ee 
In the regime {\it i}) $D_{ag}(t/t')$ remains fixed to
its initial value $D_{ag}(1)=M^2$, while $D_{st} (\tau ,\infty)$ decays to
zero. This makes Eq.~(\ref{ssplitt}) consistent with Eq.~(\ref{eqstatio}).
Conversely, in the regime {\it ii}), $D_{st}(\tau ,\infty)$ has 
already decayed
to zero and the whole time dependence is carried by $D_{ag}(t/t')$,
providing again consistence between Eqs.~(\ref{ssplitt}) and (\ref{eqaging}).

This whole pattern of behaviours of $D(t,t')$ is closely reminiscent
of what is known for the system without shear \cite{Corberi02}.
In that case it was shown that this features reflect an underlying
structure of the order parameter field which, in the
late stage of the dynamics of a quench below $T_c$, can be decomposed
into two statistically indipendent stochastic components,
$\varphi (\vec x,t)= \psi (\vec x,t) + 
\sigma (\vec x,t)$, responsible
for the stationary and aging part of the autocorrelation function
respectively,
as discussed in~\cite{Mazenko88}.
In the present case, by proceeding along the  lines of
the case $\gamma =0$, it is straightforward to show that the
same splitting of the order parameter field can be explicitly
exhibited, which accounts for the
structure of $D(t,t')$ discussed above; we refer to \cite{Corberi02,shearletter}
for further details.
In systems with  
a scalar order parameter~\cite{Corberi01} the physical
interpretation of this decomposition is the following:
given a configuration at a time  well inside the late stage scaling
regime, one can distinguish degrees of freedom $ \psi (\vec x,t)$ 
in the bulk of domains from those $ \sigma (\vec x,t)$ pertaining 
to the interfaces. The first ones are driven by thermal fluctuations
and behave locally as in equilibrium; the second ones 
retain memory of the noisy initial condition, and produce the aging behaviour.
For large-$N$, where topological defects are unstable and
domains are not well defined, the recognition of the physical 
degrees of freedom 
associated to $  \psi(\vec x,t)$ and $ \sigma(\vec x,t)$ is not
straightforward as for scalar systems. 
Nevertheless, the possibility of 
splitting of the order parameter also in this case 
indicates that the same fundamental
property is shared by systems with different $N$.

\begin{figure}[h]
    \centering
\resizebox{.7 \textwidth}{!} {\includegraphics{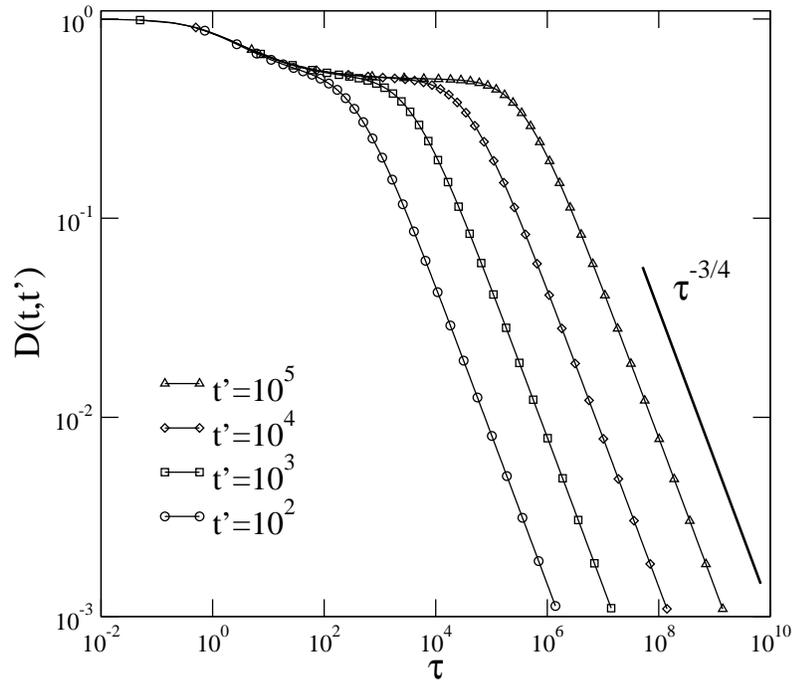}}
\vskip  - 0.8 cm
\resizebox{.7 \textwidth}{!} {\includegraphics{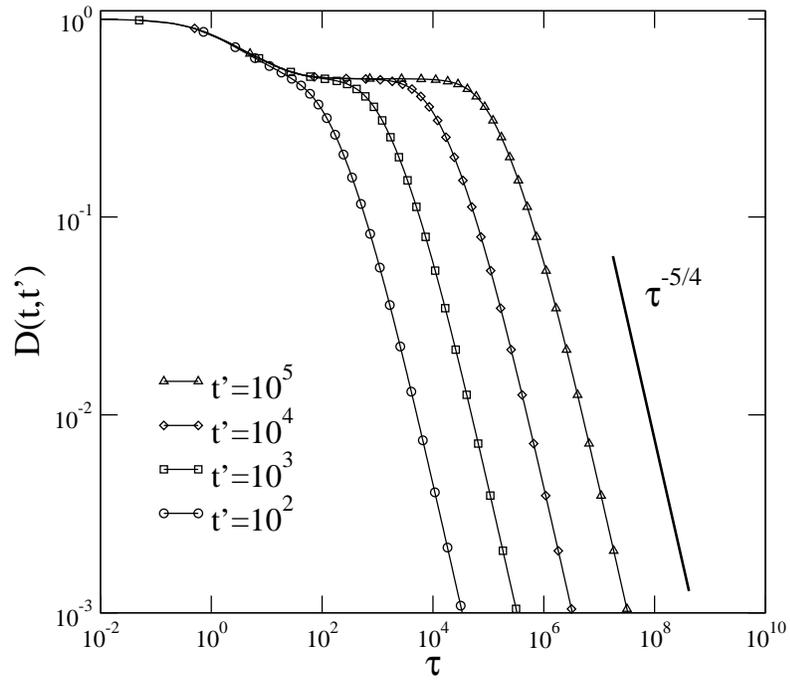} }
\caption{Autocorrelation function vs $\tau $ for
different waiting times $t'$, with
$T=T_c(\gamma)/2$, $d=3$, $\Lambda =1$.
The upper figure refers to the case without shear,
the lower one to the case with $\gamma =0.1$.}
\vskip 0.5cm
\label{aging_C}
\end{figure}

The overall behaviour of the autocorrelation function is shown in
figure \ref{aging_C}, for different $t'$.
This figure shows also a comparison with the case $\gamma =0$.
The pattern is qualitatively similar in both cases: 
Initially a fast decay
to a plateau value is observed. This is due to the vanishing 
of $D_{st}(\tau ,\infty)$ in the quasi-stationary regime while $D_{ag}(t/t')$
remains constant $D_{ag}(t/t')\simeq M^2$ (this value corresponds to
 the height of the plateau in the figure).
Notice that curves with different $t'$ collapse in this regime,
due to the stationarity of $D_{st}(\tau ,\infty)$.  
For longer times, when $t-t'\sim t'$, the autocorrelation
function departs from the plateau value, because also $D_{ag}(t/t')$
starts to fall off, and a pronounced dependence on the waiting time $t'$,
or aging, is observed. After the stationary regime is over
(around $\tau \sim 10$) also $\gamma $ plays an important role.
Actually one observes that the decay of $D_{ag}(t/t')$ is promoted by the 
shear, as expected because correlations are {\it washed out} by
the flow. Indeed, considering the case with $t'=10^5$, for example,
$D(t,t')$ decays from one to the value $10^{-3}$ in a time
$\tau $ of order $10^9$ for $\gamma =0$ and in a time of order
$3\cdot 10^7$ for $\gamma =0.1$.
For long times this effect results in the larger Fisher-Huse
exponent~(\ref{2.21b}) with respect to the undriven case,
where $\lambda=d/4$~\cite{Bray94}.

\section{Linear Response} \label{cinque}

In order to  calculate the response of the field to an
external perturbation $\vec h(\vec x, t)$ we add
to the original hamiltonian the term $\delta {\cal H} = - \int
dx^d \vec h(\vec x, t) \cdot \vec \varphi(\vec x, t)$. The Langevin
equation (\ref{eqn2}) for a generic component becomes
\begin{equation}
\frac {\partial \varphi^{\alpha} (\vec x, t)} {\partial t} + \vec \nabla
\cdot (\varphi^{\alpha}(\vec x, t) \vec v) = -     \frac {\delta {\cal
H[\vec \varphi]}}{\delta \varphi^{\alpha}(\vec x, t)} + h^{\alpha}(\vec x, t) +
\eta^{\alpha}(\vec x, t) . \label{r200}
\end{equation}
The two-time response function is defined as
\begin{equation}
        {\cal R}(\vec k,t;\vec k',t') = (2 \pi)^d
        \left . \frac {\delta  <
        \varphi^{\alpha}(\vec k, t)> }
        {\delta h^{\alpha}(-\vec k', t')}
        \right \vert _{\{h(\vec k', t')= 0  \}} \quad
        \label{r202}
\end{equation}
with  $t \geq t' $. Here,
due to rotational symmetry, a generic component of the order
parameter can  be considered and vectorial indices can  be
dropped. 
Using the properties $< \varphi^{\alpha}(\vec k,0)
> = 0 $ and $<\eta^{\alpha}(\vec k,t)> = 0 $ 
we get
\begin{equation}
{\cal R}(\vec k,t;\vec k',t') = (2 \pi)^{d} \delta \left ( \vec {\cal
K}(t-t') +\vec k' \right ) \sqrt{\frac{y(t')}{y(t)}} e^{-  
\int_0^{t-t'} du {\cal K}^2(u)} \label{r203}
\end{equation}
where the function $y(t)$ is defined in Eq.~(\ref{massren}).
The autoresponse function can be also introduced  as

\newpage
\vskip  - 3.1 cm 

\begin{eqnarray}
R(t,t') & & \equiv  \int \frac {d^d k} {(2 \pi)^d} \int \frac {d^d
k'} {(2 \pi)^d} {\cal R}(\vec k,t;\vec k',t') e^{-k^2/2\Lambda^2}
e^{-k'^2/2\Lambda^2}  \nonumber  \\ 
 & & = \sqrt{\frac{y(t')}{y(t)}} \int \frac {d^d k} {(2 \pi)^d}
 e^{-\int_0^{t-t'} du {\cal K}^2(u)}
e^{- ( k^2 + \frac {d(\gamma)}{2})/\Lambda^2} \nonumber  \\ 
 & & = \frac{2^{1-d/2}}{(4  \pi)^{d/2}}
\sqrt{\frac{y(t')}{y(t)}} \left(\frac {t-t'} {2} +
\tau _M \right)^{-d/2} \frac{1}{\sqrt{4 +
\frac{4}{3} \gamma^2 \frac {(t-t')^3} {t-t' + 2\tau _M} -
\gamma^2 \frac {(t-t')^4}{ (t-t' + 2\tau _M)^2}}}
\label{r205} \eea
where $d(\gamma) = {\cal K} ^2 (t-t')-k^2$. 

Computing   the integrated response one obtains
\begin{equation}
\chi(t,t_w) \equiv  \int_{t_w} ^t R(t,t') dt' =
\frac{2^{1-d/2}}{(4  \pi)^{d/2}}
 \frac {1}{\sqrt{y(t)}} \int _0
^{\tau}  dz \sqrt{y(t-z)} \frac {(\frac {z} {2} + \tau _M)^{-d/2}} 
{\sqrt{4 + \frac{1}{3} \gamma^2 z^2}}
\label{r207}
\end{equation}
where we have introduced the variables $z=t-t'$, $\tau = t - t_w$
and neglected the two asymptotically irrelevant
terms containing $\tau _M$  in  the square root 
in the last line of Eq.~(\ref{r205}). 
We will  evaluate this function in the
asymptotic regime $t_w \rightarrow \infty$ using the large time
behaviour of $y(t)$ given in Eqs.~(\ref{gtd2s},\ref{gtd3z}).

\vskip 1 cm
\noindent $\mathbf {T > T_c(\gamma)}$

Inserting Eqs.~(\ref{gtd2s},\ref{gtd3s})
in  Eq.~(\ref{r207}), the response function reads 
\begin{equation}
\chi(t,t_w)=\chi _{st}(\tau,\xi _\perp) = 
\frac{2}{(4  \pi)^{d/2}} \int_0
^{\tau} dz e^{- 2\xi_\perp^{-2}  z/2} \frac {(z +
2\tau _M)^{-d/2}} {\sqrt{4 + \frac{1}{3}\gamma^2 z^2}}
\label{r2077}
\end{equation}
which is a time translational invariant quantity, as expected,
because the system reaches a stationary state. 

\vskip 1 cm
\noindent $\mathbf {T= T_c(\gamma)}$

For $t_w \gg \tau _s$ one has
\be
\chi (t,t_w)=\chi _{st} (\tau, \infty )=
\frac{2}{(4  \pi)^{d/2}} \int_0
^{\tau} dz \frac {(z +
2\tau _M)^{-d/2}} {\sqrt{4 + \frac{1}{3}\gamma^2 z^2}}
\label{ccbum}
\ee
showing that also at criticality the response function is
time translational invariant.

After an integration by parts, the function 
$\chi _{st} (\tau , \infty)$ can be expressed
in $d=3$ in terms of the  $ _2 F_1 $ hypergeometric function as
\begin{eqnarray}
\chi _{st} (\tau ,\infty) & = & 
\frac{1}{(2 \pi)^{3/2}} \left[ \Lambda \left(
1/2 + (\Lambda^2 \tau +1)^{-1/2}(4+1/3\gamma^2 \tau ^2)^{-1/2}
\right ) \right ] \nonumber \\
 & - & \frac{\gamma^2}{9} \tau^{3/2}
 \mbox{ } _2F_1  \left [ 1/2, 3/4, 7/4, -\gamma^2 \tau^2/12 \right ]
\label{r20710}
\end{eqnarray}
while, in $d=2$, one gets
\begin{eqnarray}
\lefteqn{\chi _{st} (\tau ,\infty)  =
\frac {1/(2 \pi)}{\sqrt{1+1/(3 A^2)}}
\left [ \log(\Lambda^2 \tau+1) +  
\log \left (1+\sqrt{1+1/(3A^2)}\right )\right . } 
\nonumber
\\ 
 & & \left .-  \log \left (1+\sqrt{1+1/(3A^2)} \sqrt{1+1/3 (\gamma \tau/2)^2}-
\gamma \tau/(6A) \right ) \right ]  
\label{r20711}
\end{eqnarray}
In particular, for $\tau \ll \tau _s$, one has the limiting behaviour 
\be
    \chi _{st} (\tau , \infty)= \left \{
    \begin{array}{ll} \frac{1}{2\pi}\log(\Lambda^2 \tau +1)  \quad
    d=2 \\
    \frac{M_0^2}{T_c(\gamma)}\left(1- \frac{1}{\sqrt{\Lambda^2 \tau
    +1}}\right )   \quad
    d=3
    \end{array}
    \right .
    \ee
while
\be
    \chi _{st} (\tau, \infty)=\frac{M_0^2}{T_c(2\gamma)}
\label{eqchiinff}
\ee
in the opposite limit $\tau \gg \tau _s$.

\vskip 1 cm
\noindent $\mathbf {T< T_c(\gamma)}$

For  $ T < T_c(\gamma) $ the integrated response
function is given by
\begin{equation}
\chi(t,t_w)  = \frac{2}{(4  \pi)^{d/2}} \int_0^{\tau} du
\frac{(u + 1/(  \Lambda^2))^{-d/2}} {\sqrt{4 +\gamma^2
u^2/3}}
  {(1 - u/t)^{-(d+2)/4}} \qquad.
 \label{r208}
\end{equation}
The term  $(1 - u/t)^{-(d+2)/4}$ can be expanded in powers of
$u/t$ and  integrated term by term. It can be shown that the
first contribution is dominant for $t_w \gg \tau _s$
and behaves as $\chi _{st} (\tau , \infty )$
in Eq.~(\ref{ccbum}).
Then in the limit $t_w \gg \tau _s$ the response function becomes
TTI also in the low temperature phase. 

It is interesting to study the behaviour of the first correction
to this result, namely the next term appairing in the power
series in $u/t$ discussed above. 
The dependence on the two times of this quantity, 
that will be denoted as $\chi _{ag}(t,t_w)$,
can be factorized as 
$\chi _{ag}(t,t_w)=t^{-1}b(\tau)$. Here $b(\tau )$ is a TTI function that,
starting from zero at $\tau =0$, saturates to a constant value for $\tau >\tau _s$.
In the regime $\tau >\tau _s$, therefore, one has $\chi _{ag}(t,t_w)\sim t^{-1}$.
In conclusion, taking into account also this first correction,
it results that   $\chi (t,t_w)\simeq \chi _{st} (\tau , \infty )
+\chi _{ag}(t,t_w)$ and, recalling the behaviour of $\chi _{ag}(t,t_w)$,
one concludes that for large $t$ the second term is always
negligible with respect to the first, as anticipated after Eq.~(\ref{r208})

A similar structure, with a stationary and an aging part, is
also found in the case with $\gamma =0$. However, in that case, 
one finds a similar result, namely $\chi _{ag}(t,t_w)=t^{-1}b(\tau)$
and the properties of $b(\tau )$ discussed above, 
only above the upper critical dimension $d_U=4$. 
Instead, below $d_U$ it is $\chi _{ag}(t,t_w)=t_w^{-a}B(t/t_w)$, for
$\tau /t_w>1$, with $a=(d-2)/2$. Notice that this form implies that
$\chi _{ag}(t,t_w)$ does not vanish for large $t$ at the lower critical dimension 
$d=2$. A similar pattern is also found for scalar systems \cite{Corberi01}.

When shear is applied the system behaves as in the case with $\gamma =0$
and $d\geq d_U$. This is due to the fact that the mathematical structure
of the solution of the model with applied flow in dimension $d$ is
similar to the structure of the solution without shear in an
increased dimensionality $d+2$. This can be checked, for instance,
from the behaviour of the function $y(t)$ of Eqs.~(\ref{gtd2c},\ref{gtd3c})
as compared to the form without shear~\cite{Corberi02}. This phenomenon
raises the {\it effective} dimensionality at or above
$d_U=4$ for the cases $d=2,3$ considered in this paper. 

\section{Fluctuation-Dissipation relation} \label{sei}

In this Section we will
discuss the out of equilibrium generalization of the FDT, namely
the relation between $D(t,t_w)$ and $\chi (t,t_w)$ in the large
$t_w$ regime.

\vskip 1 cm
\noindent $\mathbf {T>T_c(\gamma )}$

Let us first recall that in the case without shear, 
for times $t_w$ larger than the equilibration time, 
$D(t,t_w)$ and $\chi (t,t_w)$ become TTI quantities 
$D(\tau)$ and $\chi (\tau)$.
Moreover, since the system attains an equilibrium state, standard
FDT holds
\be
T \chi (\tau) = D (\tau=0) - D(\tau) . \label{fdt}
\ee 

The effect of the shear flow is to transfer energy to the system,
and, also in the quench to a temperature above the critical point
where TTI is obeyed,
the equilibrium Gibbs state is never reached. Then the FDT does not
hold and $D(t,t_w) $ and $\chi (t,t_w)$ do not satisfy an equation analogous to
Eq.~(\ref{fdt}). The fact that $D(t,t_w)=D_{st}(\tau,\xi _\perp)$
and $\chi (t,t_w)=\chi _{st} (\tau ,\xi _\perp )$ are TTI quantities 
implies that the dependence on the two times of 
$\chi (t,t_w)$ can still be accounted
for through $D (t,t_w)$, yielding a non trivial relation
$\widetilde \chi (D)$ between $\chi(t,t_w)$ and $D(t,t_w)$.
The fluctuation-dissipation plot, namely the relation
$\widetilde \chi (D)$, is shown in Fig.~\ref{fdtsopra}.

\vskip .2cm
\begin{figure}[h]
\centering
\resizebox{.67\textwidth}{!} {\includegraphics{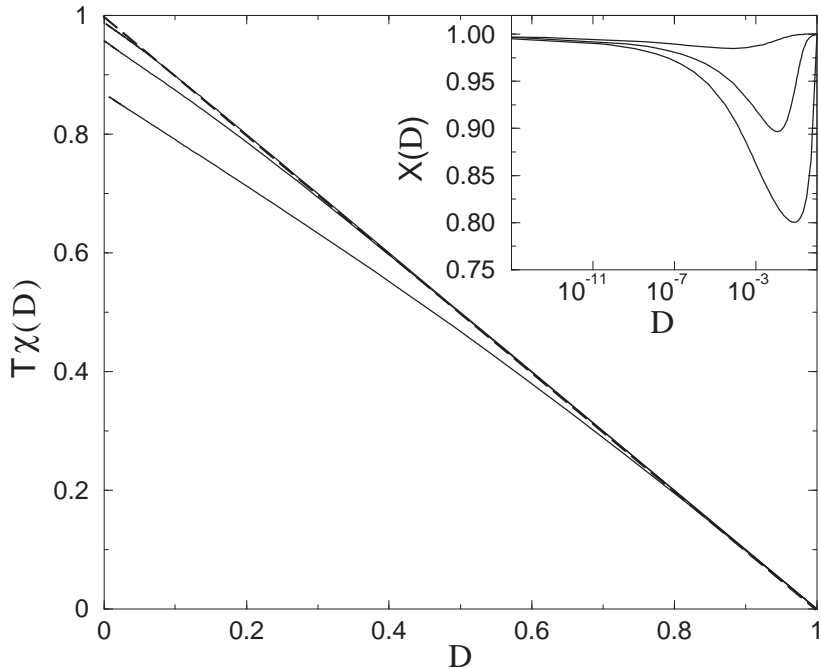} }
\caption{$T\chi (\tau )$ is plotted against 
$D(\tau )$ for a quench to a final temperature
$T=(21/20)T_c(\gamma )$, with $t_w=10^5$, $d=3$ and $\Lambda=1$. 
Solid lines correspond to 
$\gamma =10^{-2}, 10^{-1},1$, from top to bottom. The dashed line, which is
almost indistinguishable from the case $\gamma =10^{-2}$,
is the equilibrium behaviour $\gamma =0$.
The inset shows the fluctuation-dissipation ratio $X(D)$
defined in Eq.~(\ref{xxx}).}
\label{fdtsopra}
\end{figure}

The plot follows the straight line~(\ref{fdt}) for small
values of $\tau$, when $\tau \ll \tau _s$.
Recalling the discussion
of Sec.\ref{due} regarding local equilibrium
one has that the system behaves as in equilibrium up to timescales
of the order of the evolution time of the system which, in this
case,  is the flow timescale $\tau _s$. 

For larger values of $\tau $ (smaller $D$)
$\widetilde \chi (D)$ deviates 
from the equilibrium form~(\ref{fdt}) due to the effects of shear. 
In particular, for $T\gtrsim T_c(\gamma )$ its
asymptotic ($\tau =\infty $) value is  
$T\widetilde \chi (\tau ,\xi _\perp ) \approx M_0^2 T/T_c (2 \gamma)$ 
(Eq.~(\ref{eqchiinff})) 
which is less than $ M_0^2 - M^2 =  M_0^2 T/T_c(\gamma) $,
the value obtained from Eq.~(\ref{fdt}). 
From Eq.~(\ref{eqchiinff}) it is readily seen that
the relation between the response $\chi (t,t_w;2\gamma)$ of
a system subjected to a shear flow with rate $2\gamma$, 
and the autocorrelation function $D(t,t_w;\gamma)$ 
of another system with shear
rate $\gamma$, obeys Eq.~(\ref{fdt}) at large time ($\gamma \tau >>1$ ),
namely 
\be
T \chi _{st}(\tau ,\xi _\perp;2\gamma) = D_{st}(0,\xi _\perp;\gamma) 
- D_{st}(\tau ,\xi _\perp;\gamma).
\label{gam2gam}
\ee
This symmetry of the theory is independent on the  
model parameters and is therefore particularly suited
for comparison with other models and for experimental checks.
Actually, the linearization of the equation of
motion provided by the large-$N$ model is expected to provide
reliable results above the critical temperature
and we expect the relation~(\ref{gam2gam}) to be
observed also in more realistic systems.

\vskip 1 cm
\noindent $\mathbf{T=T_c(\gamma)}$

At $T=T_c(\gamma )$ the fluctuation-dissipation plot is the same
as for $T\gtrsim T_c(\gamma )$.
In the following we will focus on the 
fluctuation-dissipation ratio 
\be
X(t,t')=T\frac{R(t,t')}{\partial D(t,t')/\partial t'}.
\label{xxx}
\ee
and, in particular, on its limiting value
\be
X_\infty =\lim _{t'\to \infty} \lim _{t\to \infty } X(t,t')
\ee
In systems without drive this quantity takes the value $X_\infty =1 $ 
for equilibrium systems
while $X_\infty =0$ in the low temperature phase of coarsening
systems with a non-vanishing critical temperature \cite{Corberi02,Corberi01,Barrat98}.
For critical systems the value of $X_\infty $ has been computed for
the random walk and the Gaussian model where the non trivial value 
$X_\infty =1/2$ is found~\cite{Cugliandolo94}.
This quantity has been also computed for some coarsening systems
at the lower critical dimension $d_L$, 
such as the X-Y model in $d=2$~\cite{Cugliandolo94}
and the Ising chain~\cite{Lippiello00}, where the same result
$X_\infty =1/2$ is recovered.  
However, concerning systems quenched to the critical temperature 
above $d_L$,
the value of $X_\infty $ turns out to possess a different
value. Numerical simulations of the Ising model with heat bath dynamics 
in $d=2$ indicate that $X_\infty \sim 0.26$~\cite{Godreche00} and
the exact solution~\cite{Godreche00} of the spherical model
(that is equivalent to the large-$N$ model)
gives
\be
 X_\infty = \frac{d-2}{4} \hskip 2cm  2<d<4 
\label{bambi} 
\ee
\be
 X_\infty = \frac{1}{2} \hskip
2.3cm    d >4.
\label{bambi2} 
\ee
In~\cite{Godreche00} it is proposed that this quantity is a novel universal quantity
of non-equilibrium critical dynamics.

Under the action of a shear flow, Eqs.~(\ref{bambi2}) is
obeyed both for $d=2$ and $d=3$, as we show below. Indeed, 
concentrating on the case $d=3$ for clarity, taking
$R(t,t')$ from Eq.~(\ref{r205}) (with $y(t)=const$) in the
asymptotic limit $t-t' \gg t'$, and evaluating 
$\partial D/\partial t' $ from Eq.~(\ref{ddddhat}) one obtains
\be
 X_\infty = \frac{1}{2}.
\ee
It can be shown that the same result $X_\infty = 1/2$ applies also in
the case $d=2$. Then, with respect to the case $\gamma =0$
where Eqs.~(\ref{bambi},\ref{bambi2}) hold, the
effect of the flow is to shift by two the {\it effective} dimensionality
of the system, as already discussed in
the previous Section.

\vskip 1 cm
\noindent $\mathbf{T<T_c(\gamma)}$

Summarizing the results of the previous sections, in the regime
$t_w \gg \tau _s$ one has
\be
D(t,t_w)=D_{st} (\tau, \infty) +  D_{ag}(t/t_w) 
\label{3.1} 
\ee
\be
\chi(t,t_w)=\chi _{st} (\tau , \infty)  . 
\label{3.2} 
\ee 

Before discussing the behaviour of the flowing system it is
useful to overview the behaviour without shear that is
plotted in Fig.~\ref{fdtsotto}. In this case
a structure like~(\ref{3.1},\ref{3.2}) is also found.

\vskip .2cm
\begin{figure}[h]
\centering
\resizebox{.67\textwidth}{!} {\includegraphics{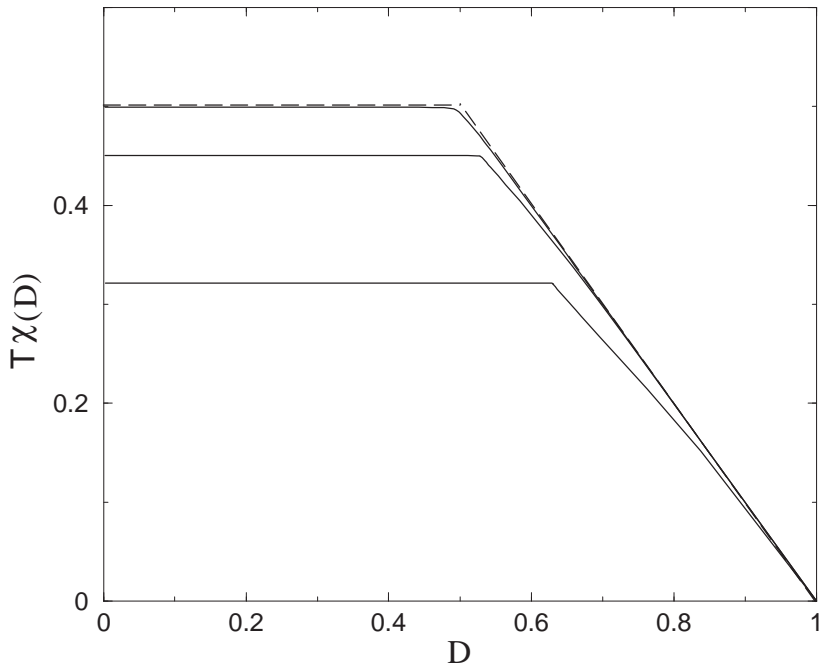} }
\caption{$T\chi (t,t_w)$ is plotted against 
$D (t,t_w)$ for a quench to a final temperature
$T=T_c(\gamma )/2$, with $t_w=10^5$, $d=3$ and $\Lambda=1$. 
Solid lines correspond to 
$\gamma = 10^{-2}, 10^{-1},1$, from top to bottom. The dashed line 
is the case with $\gamma =0$.}
\label{fdtsotto}
\end{figure}

In the short time separation regime
(see Sec.~\ref{quattro}), or stationary regime, $D_{st} (\tau ,\infty)$ 
decays from $D_{st} (0,\infty)=M_0^2-M^2$ to 
$D_{st} (\tau \simeq t_w,\infty)\simeq 0$
while $D_{ag}(t/t_w)\simeq D_{ag}(1)=M^2$ remains constant.
Therefore, in this regime $D(t,t_w)$ ranges from $M_0^2$
down to $M^2$.
$T\chi _{st}(\tau ,\infty)$ grows from zero to 
$T\chi _{st} (\tau\simeq t_w,\infty)
\simeq M_0^2-M^2$ and is related to $D_{st} (\tau , \infty)$ 
by the equilibrium FDT, Eq.~(\ref{fdt}). 
This gives rise to the straight line with 
negative slope on the right of Fig.~\ref{fdtsotto}.
Notice that this line is nothing else than the 
fluctuation-dissipation plot of a system in the equilibrium state above
$T_c$, shifted by the amount $M^2$ along the horizontal axis. 

In the large time separation sector, or aging regime,
$D_{st} (\tau ,\infty)\simeq 0$ whereas $D_{ag}(t/t_w)$ decays
from $D_{ag}(1)=M^2$ to zero for $t/t_w$ large. Hence
the aging regime corresponds to the region $0\leq D(t,t_w)\leq M^2$
of the fluctuation-dissipation plot (Fig.~\ref{fdtsotto}).
The response function stays constant $T\chi (t,t_w)=M_0^2-M^2$
because the aging part of this quantity is negligible for
large $t_w$. These behaviours account for the flat part
on the left side of Fig.~\ref{fdtsotto}.
Given this structure it is also clear that the whole time dependence
of the response can be absorbed through $D(t,t_w)$, namely
$\chi(t,t_w)=\widetilde \chi (D)$, as proposed in Ref.~\cite{Cugliandolo93}.

Next we turn to the case with shear.
In this situation, given Eqs.~(\ref{3.1},\ref{3.2}) and
the behaviour of $D_{st} (\tau ,\infty)$, $D_{ag}(t/t_w)$ and 
$\chi _{st} (\tau ,\infty)$
obtained in Secs.~\ref{quattro},\ref{cinque} 
(Eqs.~(\ref{ddddhat},\ref{eqaging},\ref{ccbum})), most of the 
considerations discussed for $\gamma =0$ still apply.
In particular 
in the short time separation regime
$D_{st} (\tau ,\infty)$ 
decays from $D_{st} (0,\infty)=M_0^2-M^2$ to zero 
while $D_{ag}(t/t_w)\simeq M^2$ keeps staying constant.
In this regime $D(t,t_w)$ ranges from $M_0^2$
down to $M^2$ and
$T\chi _{st}(\tau ,\infty)$ grows from zero to 
its limiting value $T\chi _{st} (\tau\simeq t_w,\infty)$.
However, when shear is applied, 
the relation between $\chi _{st} (\tau ,\infty)$
and $D_{st} (\tau ,\infty)$ is not the equilibrium FDT~(\ref{fdt})
but, instead, the relation $\widetilde \chi (D)$ of the case
$T\gtrsim T_c(\gamma )$ discussed above (shifted by the amount
$M^2$ along the $D$-axes, as for $\gamma =0$). 
Therefore, as discussed already for 
$T\gtrsim T_c(\gamma)$ in the stationary regime, 
one has both a region $\tau < \tau _s$ where standard
FDT~(\ref{fdt}) holds (on the right of Fig.~\ref{fdtsotto}),
and a time domain $\tau _s <\tau <t_w$ where the relation
$\widetilde \chi (D)$ deviates from the linear behaviour~(\ref{fdt}).
Finally, 
for $\tau > t_w$ (namely $D<M^2(\gamma )$),
the aging regime is explored
with a flat fluctuation-dissipation plot.  

The pattern of violation of the FDT discussed insofar for coarsening under shear
flow can be compared with the behaviour of driven mean field models for
glassy kinetics~\cite{Cugliandolo97}. 
In both cases the system without drive exhibits a phase
transition at a critical 
temperature $T_c$ characterized by an aging dynamics below $T_c$, and 
a qualitativly similar behaviour of the autocorrelation function.
However several differences occur between these systems
when the drive is switched on, some of which are revealed by the fluctuation-dissipation
relation. The most important difference is due to the fact that mean field glass
models always attain asymptotically a stationary state, regardless of the 
temperature, while, as discussed in Secs.~\ref{duestat},\ref{quattro},
the model considered in this paper exibits an aging kinetics below
a critical temperature even in the presence of shear.
Furthermore, in glassy models, the fluctuation-dissipation relation $\chi (D)$
is a broken line for every temperature and the two slopes of these lines
are associated to the existence of two
well defined temperatures in the system, the bath temperature $T$ and the effective
temperature $T_{eff}$ of the slow degrees of freedom.
A similar behaviour is found in molecular dynamics simulations of
binary Lennard-Jones mixture under shear flow~\cite{Barrat01}.
On the other hand, at least in the stationary state observed above $T_c(\gamma )$,
the only one that can be compared with the time translational invariant
states of the glassy case,
the model considered here shows a more complex relation $\chi (D)$, 
as it can be clearly observed in the inset of Fig.~\ref{fdtsotto}.  
This suggests that, despite some qualitative similarities, coarsening systems
under the action of an external drive behave quite differently from glassy
systems and that the violation of the FDT is an efficient tool to detect it.

\section{Conclusions} \label{sette}

In this paper we have studied analytically the out of equilibrium kinetics
of a solvable model for binary systems subjected to a uniform shear flow.
Besides the relevance of this subject as a first step for a better comprehension
of sheared two-component fluids, which has recently been 
by itself matter of intense theoretical and experimental
interest, this model is particularly suited for investigating off-equilibrium
systems under a general and wide perspective. Actually, 
the properties of non-equilibrium states are an important concern of
modern statistical mechanics and many efforts have been devoted 
to widen our knowledges in this field. In particular, much interest has been
payed to aging systems, namely systems which retain memory of
the time $t_w$ elapsed since they have been brought out of equilibrium, such as
glasses or spin glasses. The aging properties are usually extracted from
the behaviour of the autocorrelation function $D(t,t_w)$ 
which carries an explicit dependence on both times, up to the longest
observation scales. On the other hand, the so called driven systems,
to which external energy is pumped from the outside, generally lose
memory of  the initial condition
 and, after a microscopic time, a time translational
invariant state is entered, which, because of the external drive, is not
an equilibrium one, in the Gibbs sense. It has been argued that an important
piece of information on non-equilibrium states, both aging and stationary, is encoded into
the linear response function to an external perturbation, and, in particular,
in the so called off-equilibrium generalization of the FDT, the relation
between the response function and $D(t,t_w)$. Hopefully, this non-equilibrium
linear response theory, should bear the amount of  relevant implications and the
generality of the equilibrium case.  However, nowadays, our understanding
of this important issue is still incomplete and further investigations are necessary
in order to deepen our insight.

The model we have studied in this paper is particularly suited for
this analysis because it offers both a framework where exact calculations 
can be carried 
over and a rich phenomenology due to the presence of
both stationary and aging off-equilibrium states. 
These two behaviours are separated by a phase transition at a
critical temperature $T_c(\gamma )$
which increases with $\gamma$ and gives a transition also in $ d = 2 $.
We stress that this is a transition between
non-equilibrium states. Nevertheless, 
in this model the existence of a non-equilibrium parameter
$\gamma $ allows one to compare the properties of the transition 
with the well known equilibrium case with $\gamma =0$. In doing
that, one discovers that a wealth of similarities exist 
greatly helping the analysis of what goes on out of equilibrium.
In particular, in the low temperature phase, where the aging behaviour coexists with
the drive, the autocorrelation function can be splitted into a stationary
and an aging part, along the same lines as for $\gamma =0$.
The former describes the correlation between fast degrees of
freedom while the latter, which decays as a power law with a generalized
Fisher-Huse exponent, is due to the slow variables responsible
for the aging properties. This analysis allows 
a better comprehension of the violation of the FDT and the recognition
that a mechanism very similar to that observed without shear is at work.
This provides the basis for understanding why the fluctuation-dissipation plot
of Fig.\ref{fdtsotto}, with the flat part characteristic of
corsening systems, comes about also in sheared systems.

In aging systems without drive a definite progress in generalizing
the FDT out of equilibrium has been achieved by realizing that
certain fundamental properties of the system are encoded into the fluctuation-dissipation
relation. A first step in this direction was made by interpreting 
$T_{eff}=TX^{-1}(C)$  
as an effective temperature of the system, 
different from that of the thermal bath. 
Furthermore, it was also shown by Franz, Mezard, Parisi 
and Peliti \cite{Franz98} 
that the relation between response and autocorrelation
function bears informations on the properties of the equilibrium
state towards which the system is evolving. 
These advances qualify the fluctuation-dissipation relation as
a fundamental one also far from equilibrium, providing an important tool 
in this difficult field. 
However, a generalization of these concepts to off-equilibrium
driven systems is lacking, particularly for the low temperature phase
of the model considered here, where aging coexists with the drive. 
Further studies are therefore needed to discover if and which 
properties of the non-equilibrium state are encrypted into their
fluctuation-dissipation relation.

\section{Appendix A: The Laplace transform of f(t)}\label{appA}

In this Appendix we compute the Laplace transform of $f(t)$ defined as
\be
f^L(s)= \int_0^{\infty} e^{- t s } f(t) dt
\label{laptra}
\ee
where $f(t)$ is given in Eq.~(\ref{ftd}).
In doing so we will obtain also an explicit form for the quantity
$P(x)$ defined in Eq.~(\ref{solss}) due to the relation
\be
P(x) = 2   f^L(2   x).
\label{cucu}
\ee 
This relation can be easily derived by inserting the expression~(\ref{prior})
into Eq.~(\ref{laptra}) and comparing the result with the 
definition~(\ref{solss}).

\vskip 1cm

{\it d=2} 
\vskip 0.4cm 

We first calculate  the Laplace transform
of the function $f(t)$ of Eq.~(\ref{prior}) in $d=2$. It can be
written as
\begin{eqnarray}
f^L(s)  & = & \int_0^{\infty} e^{- t s } f(t) dt = \frac {A}{4 \pi
 } \int_0 ^{\infty} \frac{e^{ - \frac {s}{\gamma} y}}{1+Ay}
\left(4 - \frac{y^4 A^2}{(1+Ay)^2} +\frac {4}{3} \frac{y^3
A}{(1+Ay)}\right)^{-\frac{1}{2}}dy \\ & = & \frac {e^{\frac
{s}{\gamma A}}} {4 \pi  } \int_{\frac {s}{\gamma A}}
^{\infty} dz \frac {e^{-z}} {z}
 \left(4 - \frac {(z\gamma A/s -1)^4 s^2} {A^4 \gamma^2 z^2}
+\frac {4}{3} \frac{(z\gamma A/s -1)^3 s}{A^3 \gamma
z}\right)^{-\frac{1}{2}}
 \label{a1}
\nonumber
\end{eqnarray}
where  the variable $z = s (1 + A y)/(\gamma A)$ is introduced and
$A$ is defined in Eq.~(\ref{deb}). The last integral can be split
into two pieces: The first is given by
\begin{equation}
\frac {e^{\frac {s}{\gamma A}}} {8 \pi  } \int_{\frac
{s}{\gamma A}} ^{\sqrt{12} s/\gamma} dz \frac {e^{-z}}{z}
\frac{1}{\sqrt{1+h(z)}} \label{a2}
\end{equation}
where the function
\begin{equation}
h(z)= \frac{(z \gamma A/s -1)^3 s^2}{A^4 \gamma^2 z^2}
\left (\frac{z
\gamma A } {12 s} + \frac {1}{4}\right ) \label{a3}
\end{equation}
can be shown to be less than one in the integration interval.
Then the square root of Eq.~(\ref{a2}) can be expanded as a power
series and integrated term by term. The result is
\begin{equation}
\int_{\frac {s}{\gamma A}} ^{\sqrt{12} s} dz \frac {e^{-z}}{z}
\frac{1}{\sqrt{1+h(z)}}= \ln({\sqrt{12} A}) +
\sum_{n=1}^{\infty}\frac {r_n}{2n} - \sqrt{12} \frac{s}{\gamma}
\sum_{n=0}^{\infty}\frac {r_n}{2n+1}
 + {\cal O}(s^2,A^{-1})
\label{a4}
\end{equation}
with  $r_n = (-1)^n (2n -1)!!/(2n)!!$. The other contribution to
$f^L(s)$,
neglecting all terms of order ${\cal O}(s^2,A^{-1})$,
is given by
\begin{equation}
 \frac {e^{\frac {s}{\gamma A}}} {8 \pi  }
\int_{\frac {\sqrt {12} s}{\gamma }} ^{\infty} dz \frac {e^{-z}}
{z} \frac {1} {\sqrt{1 + \frac{z^2 \gamma^2}{12 s^2}}} . \label{a5}
\end{equation}
Proceeding along the same lines as for the previous integral~(\ref{a2})
and adding the result to expression~(\ref{a4}) one finds
\bea
  f^L(s)
 & = &  \frac {1}{8   \pi}
\left(\ln({\sqrt{12}A}) + 1 +  D_0 +  \sqrt{12} D_1
\frac{s}{\gamma} +\sum_{n=0}^{\infty}
\frac{12^{n+1/2}s^{2n+1}}{\gamma^{2n+1}}
\left(\prod_{k=0}^{2n}\frac{1}{k+1}\right) \ln{\frac {\sqrt{12}
s}{\gamma}}\right) \nonumber \\
  & + & {\cal O}(s^2,A^{-1})
\label{a6}
\eea where $ D_0 = \sum_{n=1}^{\infty} r_n \frac {(4n
+1)}{2n(2n+1)} =\ln 2-1 \approx -0.3068$, $ D_1 = {\cal C} - 2 - D_0
\approx -1.7296 $ with  ${\cal C} \approx 0.577216$ being the Euler's
constant. Eq.~(\ref{a6}) can be written  more compactly as
\begin{equation}
f^L(s) \approx \frac{M_0^2}{2   T_c(\gamma)} - A_2 s + B_2 s
\log s
\label{lapf2b}
\end{equation}
where  only the first term in the series  of  Eq.~(\ref{a6}) is
retained and
\begin{equation}
f^L(0)=\frac{M_0^2}{2   T_c(\gamma)}  =
 \frac {1}{8   \pi} (\ln({\sqrt{12}A}) + \ln 2 )
\label{92a}
\end{equation}
\begin{equation}
A_2 =  -  \frac { \sqrt{12}}{8   \pi}{\gamma}^{-1} (D_1 -
\ln(\sqrt{12/\gamma}))  \quad ; B_2  =  \frac { \sqrt{12}}{8
  \pi}  {\gamma}^{-1} .
\end{equation}
\vskip 1cm 
{\it d=3} 
\vskip 0.5cm 
Similarly to the case $d=2$,  starting
from  Eq.~(\ref{ftd}),
 $f^L(s)$ can be written as
\begin{eqnarray}
f^L_3(s)  & = & \frac {\sqrt{\gamma} A^{3/2}} {4 (2 \pi
 )^{3/2}} \int_0 ^{\infty} \frac{e^{ - \frac {s}{\gamma}
y}}{(1+Ay)^{3/2}} \left(4 - \frac{y^4 A^2}{(1+Ay)^2} +\frac {4}{3}
\frac{y^3 A}{(1+Ay)}\right)^{-\frac{1}{2}}dy \\ \nonumber & = &
\frac {e^{\frac {s}{\gamma A}}} {4 (2 \pi  )^{3/2}}\sqrt{s}
\int_{\frac {s}{\gamma A}} ^{\infty} dz \frac {e^{-z}} {z^{3/2}}
\left(4 - \frac {(z\gamma A/s -1)^4 s^2} {A^4 \gamma^2 z^2} +\frac
{4}{3} \frac{(z\gamma A/s -1)^3 s}{A^3 \gamma
z}\right)^{-\frac{1}{2}}
 \label{a10}
\end{eqnarray}
The last expression  can be again calculated
 splitting the integration  interval.
 A first contribution comes from the term
\begin{eqnarray}
\frac {\sqrt{s}}{2} \int_{\frac {s}{\gamma A}}
^{\sqrt{12} s/\gamma} dz \frac {e^{-z}}{z^{3/2}}
\frac{1}{\sqrt{1+h(z)}} & & = \nonumber \\ 
 \sqrt{A \gamma } + \frac {\sqrt{\gamma}} {2 (12)^{1/4}}
 \sum_{n=0}^{\infty} \frac {r_n}{2n-1/2}
& & - \frac {12^{1/4}}{2}  \frac{s}{\sqrt{\gamma}}
\sum_{n=0}^{\infty}\frac {r_n}{2n+1/2} 
 + {\cal O}(s^2,A^{-1/2}).
\label{a11}
\end{eqnarray}
Adding this contribution to the other term
\begin{eqnarray}
\frac{\sqrt{s} }{2}\int_{\frac {\sqrt {12} s}{\gamma }} ^{\infty}
dz \frac {e^{-z}} {z^{3/2}} \frac {1} {\sqrt{1 + \frac{z^2
\gamma^2}{12 s^2}}}  = \frac {\sqrt{\gamma}} {2 (12)^{1/4}}
\sum_{n=0}^{\infty}  \frac {r_n}{2n+3/2} & -  & \label{a12} \nonumber \\ 
- 12^{1/4}
\frac {s}{2 \sqrt{\gamma}}
   \sum_{n=0}^{\infty}  \frac {r_n}{2n+1/2}
 +  \frac {\Gamma(1/2)}{2} \sum _{n=0}^{\infty }\left(\prod_{k=0}^{2n+1}\frac
{2}{2k+1}\right) s^{2n+3/2} \frac {12^{n+1/2}} {\gamma^{2n +1}} &
& + {\cal O}(s^2,A^{-1/2}) 
\end{eqnarray}
one obtains
\begin{eqnarray}
f^L(s) & & =   \frac {1}{4 (2   \pi)^{3/2}} \left
[\sqrt{\gamma A} +
 \frac {\sqrt{\gamma}} {12^{1/4}} T_0
-  12^{1/4} \frac{s}{\sqrt{\gamma}}  T_1 + \frac {\Gamma (1/2)}{2}
\sum _{n=0}^\infty 
\left(\prod_{k=0}^{2n+1}\frac {2}{2k+1}\right) s^{2n+3/2} \frac
{12^{n+1/2}} {\gamma^{2n +1}} \right ]  \nonumber \\ & &
 + {\cal O}(s^2, A^{-1/2})
\label{a13}
\end{eqnarray}
where $ T_0 = \sum_{n=0}^{\infty} r_n \frac {2n
+1/2}{(2n+3/2)(2n-1/2)} \approx -0.847$, $ T_1 =
\sum_{n=0}^{\infty}r_n \frac {1}{2n+1/2} \approx 1.8541 $.
The behaviour for small $s$ is the only relevant for the
asymptotic dynamics, as discussed below Eq.~(\ref{b6}). This is given by
\begin{equation}
f^L(s)  \approx \frac{M_0^2}{2   T_c(\gamma)} - A_3 s + B_3
s^{3/2}
\label{lapf3b}
\end{equation}
with
\begin{equation}
f^L(0)=\frac{M_0^2}{2   T_c(\gamma)}  =
 \frac {1}{4 (2   \pi)^{3/2}}
(\sqrt{\gamma A}  +
 \frac {\sqrt{\gamma}} {12^{1/4}} T_0)
\label{100a}
\end{equation}
\begin{equation}
A_3 =   \frac {12^{1/4}}{4 (2   \pi)^{3/2}}
\frac{T_1}{\sqrt{\gamma}}
   \quad ; \qquad
B_3  =  \frac {\sqrt{\pi 12}}{6  (2   \pi)^{3/2}}
  {\gamma}^{-1} \quad.
\label{lapf3c}
\end{equation}
It can be shown that the coefficients $T_0, T_1$ can be also
written in terms of the $\Gamma$-function as
$T_0 = -\frac{\Gamma^2(3/4)}{\sqrt{\pi}}, T_1 = \frac{\Gamma^2(1/4)} 
{4\sqrt\pi} $.

\section {Appendix B : the function  $y(t)$}
\label{appendix B} 

We want to compute $ y(t) $, the inverse
Laplace Transformation of $y^L(s)$ of Eq.~(\ref{lapg}),  defined by
\begin{equation}
y(t) = \frac{1} {2 \pi i} \int _{\sigma  - i \infty} ^{\sigma  + i
\infty} ds \, \, y^L(s) e^{ s t} 
\label{invlap}
\end{equation}
where $\sigma$ has to be chosen in  such a way that all the poles
of $y^L(s)$ are on the left of the path of integration.

We first study the pole structure 
of $y^L(s)$
in the half-plane $x > 0 $ with $s = x + iy$.
The real part $R(x,y)$ of the denominator of $ y^L(s) $
is given by
\be
R(x,y)=x-2r -4T Re(f^L(x,y)) 
\label{b1} 
\ee 
where
\be
Re\left (f^L(x,y)\right)=\int _0^\infty dt f(t)e^{-xt}cos(yt) . 
\label{b2} 
\ee
Given that $f(t)\geq 0$ for any time, (see Eq.~(\ref{ftd})), 
for non-negative values of $x$,  $Re(f^L(x,y))$ 
is a monotone decreasing
function of both $x$ and $y$. This  implies, from Eq.~(\ref{b1}), that
$R(x,y)$ is a monotone increasing function in the  half-plane
considered. As a consequence, 
$R(x,y)$ remains positive for all $x \geq 0$ if $R(0,0)>0$. 
Since from the definition of
$T_c(\gamma)$ given in Eq.~(\ref{cicino}) one has
\be
R(0,0)>0  \mbox{,\quad  $T<T_c(\gamma) $} \label{b3} \ee
\be
R(0,0)=0 \mbox{,\quad  $T=T_c(\gamma) $} \label{b4} \ee
\be
R(0,0)<0 \mbox{,\quad  $T>T_c(\gamma) $ ,} \label{b5} \ee
for $T\leq T_c(\gamma)$, $R(x,y)$ is always positive for $
x>0 $ and $y^L(s)$ has no poles in the complex positive
half-plane.
In particular at $T=T_c(\gamma)$ there is a pole in the
origin. On the other hand,  for $T>T_c(\gamma)$, the monotonic behaviour
of $R(x,y)$ and Eq.~(\ref{b5}) imply the existence of a positive
$p$ such that $R(x,y) > R(p,0) = 0 $ for any $ x > p$. Since the
imaginary part of the denominator of  $y^L(s)$ is zero everywhere
on the real axis, one can conclude  that the largest pole is located
on the  real axis at $p$ and, in the Bromwich contour
(the vertical integration path of Fig.~\ref{contour}), 
$\sigma >p$. 

We can now calculate explicitly the integral 
of Eq.~(\ref{invlap}) in the different temperature regimes.

\vskip 1cm 
${\mathbf T<T_c(\gamma)}$

In this case there are no poles in the positive complex plane and
the Bromwich contour can coincide with the imaginary axis. The
theorem of residues  can be applied choosing a closed contour  in
the negative complex half-plane. Due to the non analytical part of 
 $f^L(s)$, there is a
branch cut in the complex plane along the negative real axis. Then we
 choose the contour $\Theta $ of Fig.~\ref{contour}
and the only non vanishing contributions come from the
Bromwich contour itself and from the integral along both sides
of the branch cut. Then, one has
\be
\int _{\Theta}ds y^L(s)e^{st}=2i\pi y(t)+ \int _0^{\infty} dx
e^{-xt}\left( y^L(s)\mid _{s=xe^{i\pi}} - y^L(s)\mid
_{s=xe^{-i\pi}} \right) . 
\label{b6} 
\ee 
Since we are interested in the
large time behaviour of $y(t)$, 
we can compute the above integrals taking only
the small $s$
behaviour of $f^L(s)$ given in Eqs.~(\ref{lapf2b},\ref{lapf3b}). We obtain
\be
\int _{\Theta}ds y^L(s)e^{st}=2i\pi y(t)- 2\pi i \frac{B_2}{M^4}
\left(T+\Delta _0 M_0^2 \right) t^{-2}  \mbox{, \quad
$d=2$ } \label{b7} \ee
\be
\int _{\Theta}ds y^L(s)e^{st}=2i\pi y(t)- 2\sqrt{\pi} i
\frac{3}{4} \frac{B_3}{M^4} \left(T+\Delta _0 M_0^2
\right) t^{-5/2} \mbox{. \quad  $d=3$ } \label{b8} \ee The above
 expressions are  equal to the sum of the residues
of all   eventual poles  inside $\Theta$.
In the negative complex plane these poles
 can  only give  contributions  exponentially decreasing with time
 which are asymptotically negligible with respect to the power law behaviour of
Eqs.~(\ref{b7}) and (\ref{b8}). Henceforth we obtain expressions~
(\ref{gtd2z},\ref{gtd3s}) in $d=2$ and $d=3$ respectively.

\begin{figure}[h]
\centering
\resizebox{.87\textwidth}{!} 
{\includegraphics{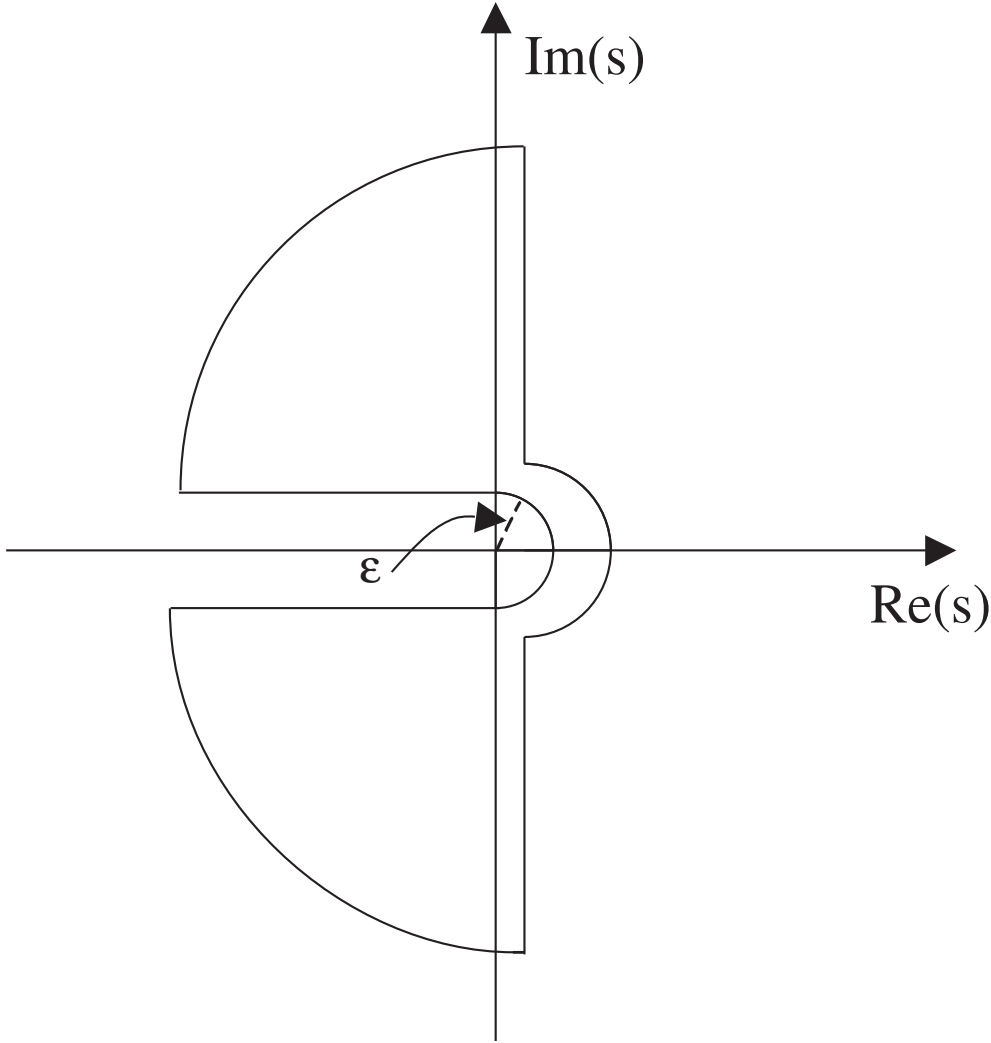} }
\caption{The closed contour adopted to compute the integral~(\ref{invlap}).}
\label{contour}
\end{figure}

\vskip 1cm
${\mathbf T=T_c(\gamma)}$

The same contour $\Theta$ of the $T<T_c(\gamma)$ case can be
chosen since   the pole located in the origin is outside $\Theta$.
The evaluation of the integrals proceeds differently in  $d=2$
and $d=3$. In $d=2$ the computation strictly follows the
one presented for $T<T_c(\gamma )$. 
The non vanishing contributions on $\Theta$ still come from
the Bromwich contour and the integrals around the cut.
One obtains
\be
\int _{\Theta}ds y^L(s)e^{st}=2i\pi y(t)- 2\pi i \frac{1}{2 B_2
T_c(\gamma)} \left(\frac{1}{2 }+ \frac{\Delta _0 M_0^2}{2
T_c(\gamma)} \right) \frac{1}{\log t} \label{b9} 
\ee 
which gives, at the
end, the result of  Eq.~(\ref{gtd2c}). In $d=3$ the contribution
coming from 
the small half-circle around the origin gives a constant
term. Since, as before,  the integral along both sides
of the cut has  a decreasing power law behaviour,
asymptotically, the contribution from the half-circle around the
origin is the dominant one   and
\be
\int _{\Theta}ds y^L(s)e^{st}=2i\pi y(t)- \pi i \frac{1+M_0^2
\Delta _0  /T_c(\gamma)}{1+4T_c(\gamma) A_3 } \label{b10}
\ee resulting in Eq.~(\ref{gtd3c}).

\vskip 1cm
${\mathbf T\gtrsim T_c(\gamma)}$

In the case of a temperature $T=T_c(\gamma)+\delta T$ with $\delta
T/T_c(\gamma)\ll 1$ the preliminary discussion about $R(x,y)$
assures the presence of a pole on the positive real axis close to
the origin.
The location $(p,0)$ of this pole is given by the solution of the equation
\be
R(p,0)=0  \label{b11} 
\ee 
that, for $p \ll 1$, as $\delta
T/T_c(\gamma)\ll 1$ implies,  reads
\be
p \log (p) = M_0^2\delta T/(T_c(\gamma)2B_2T)   \mbox{,\quad $d=2$
} 
\label{b12} 
\ee
\be
p(1/2 +2TA_3) = M_0^2\delta T/T_c(\gamma)  \mbox{,\quad
$d=3$ . } 
\label{b13} 
\ee 
At this point a similar procedure as in the case  $T=T_c(\gamma)$
can be followed giving  the
asymptotic behaviour for $y(t)$ of Eqs.~(\ref{gtd2s}) and
(\ref{gtd3s}) in $d=2$ and $d=3$, respectively.

\section{Appendix C: Asymptotic behaviour of the zero mode
of $C(\vec k, t)$ for $T  < T_c(\gamma)$} \label{appc}

\vskip 0.3cm

The value $C(\vec k=0,t_0)$ can be obtained  from
Eq.~(\ref{CahnHill}) and reads
\be
C(\vec k=0,t_0)= \frac{\Delta_0}{y(t_0)}+ \frac{2T}{y(t_0)} \int
_0 ^{t_0} y(z) dz . 
\label{1.21} 
\ee

The integral appearing in the above equation can  be splitted into
two terms:
\be
\int _0^{t_0} y(z) dz =\int _0 ^{\infty} y(z) dz - \int
_{t_0}^{\infty} y(z) dz . \label{1.12} 
\ee 
From Eq.~(\ref{lapg})
one has
\be
\int _{0}^{\infty} y(z) dz = y^L(0)=\frac{1}{2 M^2}+ \frac{M_0^2}{
M^2} \frac{\Delta_0}{2 T_c(\gamma)} \label{1.14} 
\ee 
where we have
used Eqs.~(\ref{92a},\ref{100a}), while the asymptotic
behaviour of $y(t)$ (Eqs.~(\ref{gtd2z},\ref{gtd3z})) gives
\be
\int _{t_0}^{\infty} y(z) dz =\frac{1}{(8\pi)^{d/2}}
  \frac{\sqrt {12}}{\gamma}
  \frac{T+\Delta _0 M_0^2}{M^4} \frac {2}{d}t_0^{-d/2} .
 \label{1.13}
\ee 
Putting together Eqs.~(\ref{1.21},\ref{1.14},\ref{1.13}) one finally obtains
\be
C(\vec k=0,t_0)= \frac{1}{y(t_0)}\left ( M^2  -
\frac{4T}{(8\pi)^{d/2}} \frac{\sqrt {12}}{d \gamma}
 t_0^{-d/2} \right )\frac{T+\Delta _0 M_0^2}{M^4}.
 \label{1.22}
\ee
Notice that the dependence on the initial condition, namely the
parameter $\Delta _0$, drops out for $t\gg \tau _s$ in Eq.~(\ref{1.22}),
by virtue of the expressions~(\ref{gtd2s},\ref{gtd3s}) of $y(t_0)$,
because the scaling regime is entered and the memory of the intial
state is lost.

\section{Appendix D: The autocorrelation function}

The evolution equation~(\ref{eqnd}) for the autocorrelation
function $D(t,t')$ can be integrated starting from a generic time $t_0$
chosen arbitrarely between the time of the quench ($t=0$) and the
observation time $t$. The formal solution for $D(t,t')$ then reads
\begin{equation}
D(t,t') = D_{T}(t,t') +D_{0}(t,t') 
\label{2.2}
\end{equation}
where
\begin{equation}
D_{T}(t,t') =
 \frac{2   T}{\sqrt{y(t') y(t)}}
 \int \frac {d^d k} {(2 \pi)^d}
e^{- ( k^2 + d(\gamma)/2)/\Lambda^2}
  \int_{t_0}^{t'}  dz
e^{-2   \int _{0}^{t-z} {\cal K}^2(u)du} e^{   \int
_{0}^{t-t'} {\cal K}^2(u)du}   y(z) 
\label{2.4}
\end{equation}
and
\begin{equation}
D_{0}(t,t') = \frac{y(t_0)} {\sqrt{y(t') y(t)}}
 \int \frac {d^d k} {(2 \pi)^d}  C\left ({\cal K}(t-t_0),t_0 \right )
e^{- ( k^2 + d(\gamma)/2)/\Lambda^2}
  {e^{- 2   \int _0 ^{t-t_0} {\cal K}^2(u) du}
e^{   \int _0 ^{\tau} {\cal K}^2(u) du}} \label{2.3}
\end{equation}
with $ d(\gamma) =  {\cal K}^2(t-t') -k^2$.

We will take the time $t_0$ large enough
($t_0 \gg \tau _s$) that  $y(t)$ is given by
Eqs.~(\ref{gtd2s},\ref{gtd3s}). Furthermore it can be shown that,
due to the condition~(\ref{deb}), the function  $ d(\gamma) $
can be neglected in Eqs.~(\ref{2.3},\ref{2.4}).

We will first  consider the function  $D_T(t,t')$. Carrying out the integral
one obtains
        \begin{eqnarray}
            \lefteqn{D_{T}(t,t')= \frac {4T}{(8\pi )^{d/2}\sqrt {y(t) y(t')}}
            \times }  \label{2.7} \\ \nonumber
            & &  \int _{t_0}^{t'}
            y(z) (t_m-z +\tau _M)^{-d/2} \frac {1}{\sqrt{4- \gamma ^2
            \frac {((t-z)^2-\tau^2/2)^2}{ (t_m-z+\tau _M)^2} +
            \frac{4}{3}\gamma ^2 \frac {((t-z)^3-\tau^3/2)}{(t_m-z
            +\tau _M)}}} dz \
            \end{eqnarray}
        with $  \tau = t-t'  $ and $
        t_m=(t+t')/2$.
The  role of the cutoff $\Lambda$  is crucial in the factor 
$(t_m-z +\tau _M)^{-d/2}$
in order to avoid divergences at $z=t_m$; however, 
for large $t_m >\tau _s$, one finds that it can be neglected in the two 
terms under the square root. Then,
introducing the variable $u=1-z/t_m$, one has
\be
    D_{T}(t,t')= \frac {4T}{(8\pi )^{d/2}}\frac{t_m^{-d/2}}{\gamma}
    \int _{\tau /(2t_m)}^{1-t_0/t_m}
    \rho (u,\tau,t_m) \Theta (u,\tau,t_m) du
    \label{53} 
\ee 
where
\be
    \rho (u,\tau,t_m) = \frac{(u +1/(2\Lambda^2t_m))^{-d/2}}
    {\sqrt{\frac{4}{\gamma^2t_m^2} +\frac{1}{3} u^2 + \frac{1}{2}
    (\frac{\tau}{t_m}) ^2 \left (1-\frac{(\tau/t_m)^2}{8u^2} \right )
    }}  \label{54} \ee
and
\be
            \Theta(u,\tau,t_m)
        = \left \{ \begin{array}{ll}
        e^{-2 \xi_\perp ^{-2}  t_m u}
        & \mbox{; for $T>T_c(\gamma)$} \\ \nonumber
          1 &\mbox{; for $T=T_c(\gamma)$, $d=3$} \\ \nonumber
         \frac {\sqrt {\log(t_m+\tau/2) \log(t_m-\tau/2)}}{\log[t_m(1-u)]}
          &  \mbox{; for $T=T_c(\gamma)$, $d=2$} \\ \nonumber
         \left( \frac{1-u}{\sqrt{1-(\tau/2t_m)^2}} \right )^{-\frac{d+2}{2}}
        & \mbox{; for $T<T_c(\gamma)$ .}
        \end{array}
        \right .
        \label{55}
    \ee

Turning now to the function $D_0(t,t')$, in the
limit $t'/t_0 \gg 1$, due to the presence of the exponential
factors  in Eq.~(\ref{2.3}),  $D_0(t,t')$ can be evaluated by
approximating the one-time correlation function 
$C\left ( {\cal K}(t-t_0),t_0 \right )$ 
at the time $t_0$  
entering Eq.~(\ref{2.4}), 
whose expression is given in 
Eq.~(\ref{CahnHill}), 
with its value at  $\vec k =
0 $. Then one has
\begin{eqnarray}
        \lefteqn{D_{0}(t,t')=}  \label{2.6} \\ \nonumber
            & & \frac {C(\vec k = 0 ,
            t_0)y(t_0)} {\sqrt {y(t) y(t')}}
            (t-\tau/2+1/2\Lambda^2)^{-d/2}
            \frac {2}{\sqrt{4- \gamma ^2 \frac {(t^2-\tau^2/2)^2}
            {(t-\tau/2+1/(2\Lambda^2))^2} + \frac{4}{3}\gamma ^2 \frac
            {(t^3-\tau^3/2)}{(t-\tau/2+1/2\Lambda^2) }}} .
\label{2.61111} 
\end{eqnarray} 
From this point on one must distinguish different temperature ranges.

\vskip 1cm
\noindent $\mathbf {T> T_c(\gamma )}$

Starting with $D_T(t,t')$, from
Eqs.(\ref{53},\ref{54},\ref{55}), one has
\be
D_{T}(t,t')= \frac {4T}{(8\pi )^{d/2}} \int _{\tau/2}^{t_m-t_0}
    e^{-2 \xi_\perp^{-2}   y}
    (y+1/2\Lambda^2)^{-d/2} \frac {1}{\sqrt{4 +\frac{1}{3}\gamma ^2 y^2
    + \frac{\gamma ^2}{2} \tau ^2 \left (1-\frac{\tau^2}{8y^2} \right
    ) }}dy.
    \label{52d} \ee
In the limit $t_m \to
\infty$, $D_T(t_m,\tau)$ becomes a TTI quantity
\begin{equation}
D_{st}(\tau,\xi_\perp) = \frac {4T}{(8\pi )^{d/2}} \int
_{\tau/2}^\infty
    e^{-2 \xi_\perp^{-2}   y}
    (y+1/2\Lambda^2)^{-d/2} \frac {1}{\sqrt{4 +\frac{1}{3}\gamma ^2 y^2
    + \frac{\gamma ^2}{2} \tau ^2 \left (1-\frac{\tau^2}{8y^2} \right
    ) }}dy.
\label{ddhat}
\ee

With respect to $D_0(t,t')$, by inserting the asymptotic behaviours
of $y(t)$ ( Eqs.~(\ref{gtd2s},\ref{gtd3s})) in Eq.~(\ref{2.6}),  one
immediately obtains
\be
    D_0(t,t') \sim e^{-2  \xi_\perp^{-2}(t+t')/2}
\label{152a}
\ee

From expressions~(\ref{ddhat},\ref{152a}) it is clear that, in the limit
$t' \gg \xi _\perp ^2$, $D_0(t,t')$ is negligible with respect to
$D_T(t,t')$ so that 
\be
D(t,t')\simeq D_{st}(\tau, \xi _\perp)
\label{riass1}
\ee
and $D_{st}(\tau, \xi _\perp)$ is given in Eq.~(\ref{ddhat}).  

\vskip 1cm
\noindent $\mathbf {T= T_c(\gamma )}$

In this case the same results~(\ref{52d},\ref{ddhat}) are found for $D_T(t,t')$
but with  $\xi_\perp^{-1}=0$. In particular the large $t_m$ behaviour of
$D(t,t')$ is given by
\begin{equation}
D_{st}(\tau,\infty) = \frac {4T}{(8\pi )^{d/2}} \int
_{\tau/2}^\infty
    (y+1/2\Lambda^2)^{-d/2} \frac {1}{\sqrt{4 +\frac{1}{3}\gamma ^2 y^2
    + \frac{\gamma ^2}{2} \tau ^2 \left (1-\frac{\tau^2}{8y^2} \right
    ) }}dy
\label{ddhattc}
\ee
with corrections of order $t_m^{-d/2}$.

Regarding $D_0(t,t')$, from Eqs.~(\ref{gtd2s},\ref{gtd3s}) one has
\be
    D_0(t,t') \sim \left (\frac {t+t'}{2} \right )^{-(d+2)/2}
\label{152b}
\ee
with logarithmic corrections in $d=2$.

Comparing Eq.~(\ref{ddhattc}) with Eq.~(\ref{152b}) one can show that again, 
in the limit $t' \gg \tau _s$, $D_0(t,t')$ is negligible with respect to
$D_T(t,t')$, as for $T>T_c(\gamma )$.
Therefore 
\be
D(t,t')\simeq D_{st}(\tau, \infty).
\label{riass2}
\ee
The expression for $D_{st}(\tau, \infty)$ is Eq.~(\ref{ddhattc}).  

\vskip 1cm
\noindent $\mathbf {T<T_c(\gamma )}$

For $T<T_c(\gamma)$ we study the behaviour of $D_T(t,t')$ separately in
the temporal regimes
$\tau/t_m \ll 1 $ ({\it quasi-stationary regime}) and
$\tau/t_m \gg  2/3$ $(\tau/t' \gg 1)$ ({\it aging regime}). 
In the quasi-stationary regime, the function $\rho (u, \tau ,t_m)$ 
of Eq.~(\ref{54}) is of
order $t_m^{(d+2)/2}$ for $u \ll 1$ and decreases to zero as 
$\rho (u,\tau ,t_m)
\sim u^{-(d+2)/2}$ when $u>1/\gamma t_m$. In the same regime,
$\Theta \simeq 1$ when  $u \ll 1$ while it diverges like
$(1-u)^{-(d+2)/2}$ for $u \to 1-t_0/t_m$. In the limits
$\gamma t_m \to \infty$ and $t_0/t_m \to 0$, the leading contributions
to the integral of Eq.~(\ref{53}) coming from
the singularities of
the functions $\rho(u,\tau,t_m)$ for $u \to 0$ and of $\Theta
(u,\tau,t_m)$ for $u \to 1-t_0/t_m$ give
    \bea
    & & D_T(t_m,\tau)  = \label{64} \\ \nonumber
    & &  \frac {4T}{(8\pi )^{d/2}}\frac{t_m^{-d/2}}{\gamma}
    \left [ \int_{\tau/(2t_m)}^\infty du \rho(u,\tau,t_m)
    + \sqrt3 \int_0^{1-t_0/(2t_m)} du \Theta(u,\tau,t_m) \right ]
    + {\cal O}(t_m^{-d/2}) = \\ \nonumber
    & & D_{st}(\tau ,\infty)
    + \frac{4\sqrt{12}T}{(8\pi )^{d/2}d \gamma}t_0^{-d/2}
    + {\cal O}(t_m^{-d/2}) .
    \eea
\vskip 0.5cm 
Similar considerations applied to the aging regime imply
 \bea
 & &  D_T(t_m,\tau)  = \label{66b} \\ \nonumber
    & &  \frac {4T}{(8\pi )^{d/2}}\frac{t_m^{-d/2}}{\gamma}
    \frac{\sqrt3}{\sqrt{1+\frac{3}{2}(\frac{\tau}{t_m})^2(1-\frac{1}{8}
    (\frac{\tau}{t_m})^2)}}
     \int_0^{1-t_0/(2t_m)} du \Theta(u,\tau,t_m)
    + {\cal O}(t_m^{-d/2}) = \\ \nonumber
    & & \frac {4T}{(8\pi
    )^{d/2}} \frac{\sqrt{12}}{d\gamma}t_0^{-d/2}
    \left ( \frac{t}{t'} \right )^{-\frac{d+2}{4} }
    (1+t'/t)^{-\frac{d+2}{2}}
    \frac {2^\frac{d+2}{2}}{ \sqrt{ 4 \frac {2-(1-t'/t)^3}{1+t'/t}- 3
    \frac {(2-(1-t'/t)^2)^2}{(1+t'/t)^2} }}
     + {\cal O}(t_m^{-d/2}).
    \eea

For $D_0(t,t')$ one finds
\bea
    \lefteqn{D_{0}(t,t') =}   \label{66} \\ \nonumber
    & & \left (M^2- \frac {4T}{(8\pi
    )^{d/2}} \frac{\sqrt{12}}{d\gamma}t_0^{-d/2} \right )
    \left ( \frac{t}{t'} \right )^{-\frac{d+2}{4} }
    (1+t'/t)^{-\frac{d+2}{2}}
    \frac {2^\frac{d+2}{2}}{ \sqrt{ 4 \frac {2-(1-t'/t)^3}{1+t'/t}- 3
    \frac {(2-(1-t'/t)^2)^2}{(1+t'/t)^2} }}  
\eea
where the expression of $C(\vec k=0,t_0)$
obtained in appendix C has been used
and the limit $t'\gg \tau _s$ is assumed. 

Summarizing, summing up $D_0(t,t')$ (Eq.~(\ref{66})) with $D_T(t,t')$ 
(Eq.~(\ref{64})), in the quasi-stationary regime one obtains
\be
D(t,t')\simeq D_{st}(\tau,\infty)+M^2
\label{riass3}
\ee
with $D_{st}(\tau,\infty)$ given in Eq.~(\ref{ddhattc})
while, in the aging regime, from Eqs.~(\ref{66},\ref{66b}) one has
\be
  D(t,t')=D_{ag}(t/t')= M^2
    \left ( \frac{t}{t'} \right )^{-\frac{d+2}{4} }
    (1+t'/t)^{-\frac{d+2}{2}}
    \frac {2^\frac{d+2}{2}}{ \sqrt{ 4 \frac {2-(1-t'/t)^3}{1+t'/t}- 3
    \frac {(2-(1-t'/t)^2)^2}{(1+t'/t)^2} }} . 
\label{riass4}
\ee
thus defining the quantity $D_{ag}(t/t')$.

\vskip 1cm

\acknowledgments 
We acknowledge support by MURST(PRIN 2000).
F.C. is grateful to M.Cirillo, R.Del Sole and
M.Palummo for hospitality in the University of Rome. 

\dag corberi@na.infn.it ; *gonnella@ba.infn.it

\ddag lippiello@sa.infn.it ; \S zannetti@na.infn.it


\begin{references}

\bibitem{Hohenberg89}
P.C.~Hohenberg and B.I.~Shraiman, Physica D {\bf 37}, 109 (1989).

\bibitem{Cugliandolo93}
L.F.~Cugliandolo and J.~Kurchan, Phys.Rev.Lett. {\bf 71}, 
173 (1993); Philos.Mag. {\bf 71}, 501 (1995); 
J.Phys. A {\bf 27}, 5749 (1994).

\bibitem{Cugliandolo97}
L.F.~Cugliandolo, J.~Kurchan, and L.~Peliti
Phys. Rev. E {\bf 55}, 3898 (1997);
L.~Berthier, J.-L.~Barrat and J.~Kurchan, Phys. Rev. E 
{\bf 61}, 5464 (2000).

\bibitem{Fielding02}
S.~Fielding and P.~Sollich, Phys. Rev. Lett. {\bf 88}, 050603 (2002);
cond-mat/0209645.

\bibitem{Schmittman95}
B.~Schmittmann and R.K.P.~Zia, ``Statistical Mechanics of Driven
Diffusive Systems'' in ``Phase Transitions and Critical
Phenomena'', Vol. 17, eds. C. Domb and J.L. Lebowitz, (Academic
1995).

\bibitem{Gallavotti95}
G.~Gallavotti and E.G.D~Cohen, Phys. Rev. Lett. {\bf 74}, 2694 (1995);
J. Stat. Phys. {\bf 80}, 931 (1995).

\bibitem{Larson99}
See, e.g., R.G.~Larson, {\it The structure and Rheology of Complex
fluids} (Oxford University Press, New York, 1999).

\bibitem{Onuki97}
For a review, see A.~Onuki, J. Phys.: Condens. Matter {\bf 9},
6119 (1997).

\bibitem{Corberi99}
F.~Corberi, G.~Gonnella, and A.~Lamura, Phys. Rev. Lett. {\bf 83},
4057 (1999).

\bibitem {Corberi98}
F.~Corberi, G.~Gonnella, and A.~Lamura, Phys. Rev. Lett. {\bf 81},
3852 (1998).

\bibitem{braz}
F.~Corberi, G.~Gonnella and A.~Lamura, Phys. Rev. E {\bf 66}, 016114
(2002).

\bibitem{Bray94}
A.J.~Bray, Adv. in Phys. {\bf 43}, 357 (1994).

\bibitem{Cavagna00}
A.J.~Bray and A.~Cavagna, J. Phys A: Math. Gen. {\bf 33}, L305
(2000).

\bibitem{Coniglio94}
A.~Coniglio, P.~Ruggiero, and M.~Zannetti, Phys. Rev. E {\bf 50},
1046 (1994).

\bibitem{Onuki79}
 A.~Onuki and K.~Kawasaki, Ann. Phys. {\bf 121}, 456 (1979).

\bibitem{Kreuzer81}
H.J.~Kreuzer, ``Nonequilibrium Thermodynamics and its Statistical
Foundations'', eds. H.~Fr\"olich, P.B.~Hirsch and N.F.~Mott,
(Oxford University Press 1981).

\bibitem{Pellicoro}

G.~Gonnella and M.~Pellicoro, J. Phys. A, 7043 (2000).
C.~Chan and L. Lin, Europhys. Lett. {\bf 11}, 13 (1990).

\bibitem{Corberi02}
F.~Corberi, E.~Lippiello, and M.~Zannetti, 
Phys. Rev. E {\bf 65}, 046136 (2002).

\bibitem{Mazenko88}
G.F.~Mazenko, O.T.~Valls and M.Zannetti, Phys.Rev. B {\bf 38},
520 (1988);
S.~Franz and M.A.~Virasoro, J.Phys. A {\bf33}, 891 (2000). 

\bibitem{shearletter}
F.~Corberi, G.~Gonnella, E.~Lippiello and M.~Zannetti,
cond-mat/0205627, to appear on Europhys. Lett.

\bibitem{Corberi01}
F.~Corberi, E.~Lippiello and M.~Zannetti, Phys.Rev. E 
{\bf 63}, 061506 (2001); Eur.Phys.J. B.
{\bf 24}, 359 (2001).

\bibitem{Barrat98}
A.~Barrat, Phys.Rev.E {\bf 57}, 3629 (1998).

\bibitem{Cugliandolo94}
L.F.~Cugliandolo, J.~Kurchan and G.~Parisi, J.Phys.I France 
{\bf 4}, 1641 (1994).

\bibitem{Lippiello00}
C.Godreche and J.M.Luck, J.Phys. A {\bf 33}, 1151 (2000);
E.Lippiello and M.Zannetti, Phys.Rev. E {\bf 61}, 3369 (2000).

\bibitem{Godreche00}
C.~Godreche and J.M.~Luck,  J. Phys A {\bf 33}, 9141 (2000).

\bibitem{Barrat01}
J.-L.~Barrat and L.~Berthier, Phys. Rev. E {\bf 63},
012503 (2001); cond-mat/0111312.

\bibitem{Franz98}
S.~Franz, M.~Mezard, G.~Parisi and L.~Peliti, Phys.Rev.Lett. {\bf 81}, 
1758 (1998); J.Stat.Phys. {\bf 97}, 459 (1998).

\end{references}
\enddocument